\documentclass{99-Styles/MICE}
\usepackage{lineno}
\usepackage{bibentry}
\nobibliography*
\usepackage{times}
\usepackage{bm}         % bold maths
\usepackage{amsmath,amssymb,bm,slashed} % need for subequations
\usepackage{amssymb}    % extra math symbols
\usepackage{graphicx}   % need for figures
\usepackage{verbatim}   % useful for program listings
\usepackage{color}      % use if color is used in text
\usepackage{subfigure}  % use for side-by-side figures
\usepackage{hyperref}   % use for hypertext links, including those to external documents and URLs
\usepackage{enumerate}
\usepackage{pdflscape}

\usepackage{fmtcount}   % Allow counters to be printed with padding

\usepackage[square,comma,numbers,sort&compress]{natbib}

\usepackage{authblk}	% for author list with affiliations
\usepackage{multirow}

\modulolinenumbers[5]

\begin{document}

\thispagestyle{empty}
 
\begin{tabular}{p{0.175\textwidth} p{0.5\textwidth} p{0.225\textwidth}}
  \hspace{-0.8cm}\leftline{\today}                                      &
  \centering{Performance of the MICE diagnostic system}                &
%  \rightline{Draft 6.6} 
  \rightline{RAL-P-2021-001} 

\end{tabular}
\vspace{-1.0cm}\\
\rule{\textwidth}{0.43pt}

\begin{center}
  {\bf
    {\LARGE Performance of the MICE diagnostic system} \\
  }
  \vspace{0.2cm}
  The MICE Collaboration\\
  \vspace{-0.0cm}
\end{center}

\makeatletter

% Quantum mechanical notation
\newcommand{\bra}[1]{\ensuremath{\langle #1 |}}   % Bra vector
\newcommand{\ket}[1]{\ensuremath{| #1 \rangle}}   % Ket vector
\newcommand{\bigbra}[1]{\ensuremath{\big\langle #1 \big|}}   % Bra vector
\newcommand{\bigket}[1]{\ensuremath{\big| #1 \big\rangle}}   % Ket vector
\newcommand{\amp}[3]{\ensuremath{\left\langle #1 \,\left|\, #2%
                     \,\right|\, #3 \right\rangle}}  % QM amplitude
\newcommand{\sprod}[2]{\ensuremath{\left\langle #1 |%
                     #2 \right\rangle}}  % QM scalar product
\newcommand{\ev}[1]{\ensuremath{\left\langle #1 %
                     \right\rangle}} % Expectation value
\newcommand{\ds}[1]{\ensuremath{\! \frac{d^3#1}{(2\pi)^3 %
                     \sqrt{2 E_\vec{#1}}} \,}} % Spatial integral
\newcommand{\dst}[1]{\ensuremath{\! %
                     \frac{d^4#1}{(2\pi)^4} \,}} % Space-time integral
\newcommand{\tr}{\text{tr}}
\newcommand{\sgn}{\text{sgn}}
\newcommand{\diag}{\text{diag}}
\newcommand{\BR}{\text{BR}}
\newcommand{\gsim}      {\mbox{\raisebox{-0.4ex}{$\;\stackrel{>}{\scriptstyle \sim}\;$}}}
\newcommand{\lsim}      {\mbox{\raisebox{-0.4ex}{$\;\stackrel{<}{\scriptstyle \sim}\;$}}}

% Miscellaneous commands
\renewcommand{\vec}[1]{{\mathbf{#1}}}
\renewcommand{\Re}{{\text{Re}}}
\renewcommand{\Im}{{\text{Im}}}
\newcommand{\iso}[2]{{\ensuremath{{}^{#2}}\ensuremath{\rm #1}}}
\newcommand{\eps}{{\ensuremath{\epsilon}}}
\newcommand{\draftnote}[1]{{\bf\color{red} \MakeUppercase{#1}}}
\newcommand{\panm}[1]{{\color{blue} #1}}
\providecommand{\abs}[1]{\lvert#1\rvert}
\providecommand{\norm}[1]{\lVert#1\rVert}

\def\parenbar{\mathpalette\p@renb@r}
\def\p@renb@r#1#2{\vbox{%
  \ifx#1\scriptscriptstyle \dimen@.7em\dimen@ii.2em\else
  \ifx#1\scriptstyle \dimen@.8em\dimen@ii.25em\else
  \dimen@1em\dimen@ii.4em\fi\fi \offinterlineskip
  \ialign{\hfill##\hfill\cr
    \vbox{\hrule width\dimen@ii}\cr
    \noalign{\vskip-.3ex}%
    \hbox to\dimen@{$\mathchar300\hfil\mathchar301$}\cr
    \noalign{\vskip-.3ex}%
    $#1#2$\cr}}}

%
%.. Sanjib's commands:
\providecommand{\anmne}{\mbox{$\bar\nu_{\mu} \rightarrow \bar\nu_e$}} 
\providecommand{\nmne}{\mbox{$\nu_{\mu}\rightarrow\nu_e$}} 
\providecommand{\anm}{\mbox{$\bar\nu_\mu$}} 
\providecommand{\nm}{\mbox{$\nu_\mu$}}
\providecommand{\nue}{\mbox{$\nu_e$}} 
\providecommand{\ane}{\mbox{$\bar\nu_e$}} 
\providecommand{\enu}{\mbox{$E_\nu$}}
\providecommand{\piz}{\mbox{$\pi^0 $}}
\providecommand{\pip}{\mbox{$\pi^+$}} 
\providecommand{\pim}{\mbox{$\pi^-$}}

\parindent 10pt
\pagestyle{plain}
\pagenumbering{arabic}                   
\setcounter{page}{1}

\begin{quotation}
\begin{center}
\textbf{Abstract}
\end{center}

\noindent
Muon beams of low emittance provide the basis for the intense,
well-characterised neutrino beams of a neutrino factory and for
multi-TeV lepton-antilepton collisions at a muon collider.
The international Muon Ionization Cooling Experiment (MICE) has
demonstrated the principle of ionization cooling, the technique by
which it is proposed to reduce the phase-space volume occupied by the
muon beam at such facilities. 
This paper documents the performance of the detectors used in MICE
to measure the muon-beam parameters, and the physical properties of the liquid hydrogen energy absorber during running.

\end{quotation}

%\tableofcontents

\graphicspath{{01-Introduction/Figures/}}

\section{Introduction}
\label{Sect:Intro}

Stored muon beams have been proposed as the basis of a facility
capable of delivering lepton-antilepton collisions at very high
energy~\cite{Neuffer:1994bt,Palmer:2014nza} and as the source of
uniquely well-characterised neutrino 
beams~\cite{Geer:1998PhRvD..57.6989G,Bandyopadhyay:2007kx,Apollonio:2002en}.
In the majority of designs for such facilities the muons are produced
from the decay of pions created when an intense proton beam strikes a
target.
The phase-space volume occupied by the tertiary muon beam must be
reduced (cooled) before the beam is accelerated and subsequently injected
into a storage ring.
The times taken to cool the beam using techniques that are presently in
use at particle accelerators (synchrotron-radiation cooling
\cite{2012acph.book.....L}, laser
cooling~\cite{PhysRevLett.64.2901,PhysRevLett.67.1238,doi:10.1063/1.329218},
stochastic cooling~\cite{Marriner:2003mn}, electron
cooling~\cite{1063-7869-43-5-R01} and frictional cooling~\cite{PhysRevLett.125.164802})
are long when compared with the lifetime of the muon.
%The use of such techniques would therefore lead to unacceptably large
%losses through decay.
Ionization cooling~\cite{cooling_methods,Neuffer:1983jr}, in which a
muon beam is passed through a material (the absorber) where it
loses energy, and is then re-accelerated, occurs on a timescale short
compared with the muon lifetime.
Ionization cooling is therefore the only technique available to cool the muon beam at a neutrino factory or muon collider.
The international Muon Ionization Cooling Experiment (MICE)
provided the proof-of-principle demonstration of the
ionization-cooling technique~\cite{Bogomilov:2019kfj}.

\begin{figure}[htb!]
  \begin{center}
    \includegraphics[width=1.0\columnwidth]{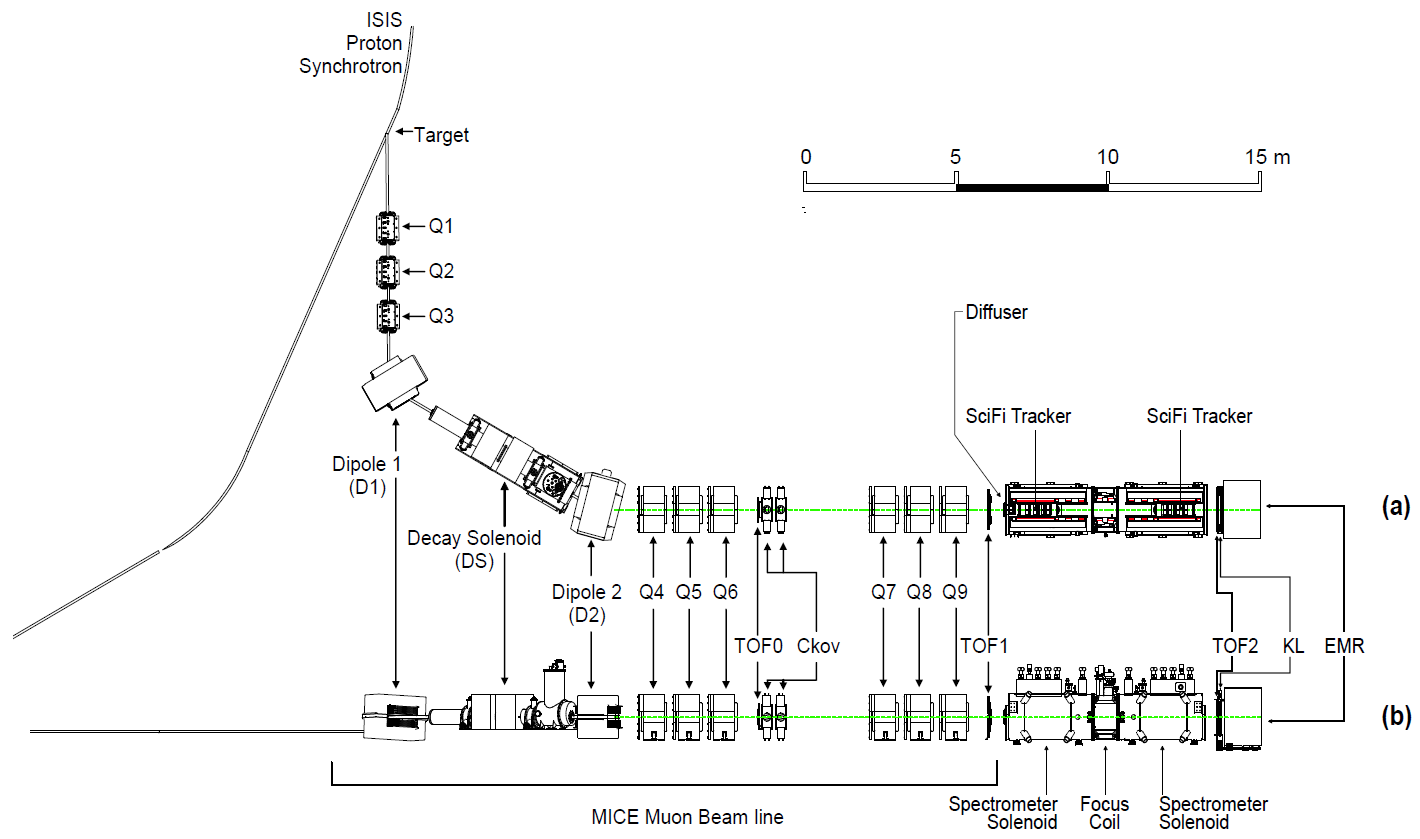}
    \caption{
      MICE, top (a) and side (b) views, showing the full beam line
      starting from the target position on the proton synchrotron with the quadrupoles and dipoles (Q1 to Q9, D1, D2), the
      Decay Solenoid, and instrumented magnetic channel elements
      (including the trackers upstream, TKU, and downstream, TKD, of the cooling
      channel, placed inside superconducting solenoids, respectively SSU and SSD) with all the
      other PID detectors (three TOF stations, two Ckov detectors, KL and
      the EMR).
      The cooling cell, defined to be the liquid hydrogen absorber
      vessel inside the focus coil (FC), is shown in figure~\ref{Fig:AbsorberVessel:Diag}.
    }
    \label{fig:BL}
  \end{center}
\end{figure}

MICE operated at the ISIS Neutron and Muon Source at the STFC
Rutherford Appleton Laboratory.
% from 2008 to 2018.  
The ISIS synchrotron accelerates pulses of protons to a kinetic energy of 800\,MeV at 50\,Hz.
For MICE operation, a titanium target was dipped
%nearly once per second
into the halo of the proton beam at 0.78\,Hz. 
Pions created in the interaction of the beam and target were captured in a quadrupole triplet
(see figure~\ref{fig:BL}).
A beam line composed of dipole, solenoid, and quadrupole
magnets captured muons produced through pion decay and transported the
resulting muon beam to the MICE apparatus.
The momentum of the muon beam was determined by the settings of the two dipole magnets D1 and D2.
Beams having muon central momenta between 140\,MeV/$c$ and 240\,MeV/$c$ were used for ionisation cooling studies.
The emittance of the beam injected into the experiment was tuned using
a set of adjustable diffusers, some made of tungsten and some of brass.
The cooling cell was composed of a liquid hydrogen or lithium hydride
absorber placed inside a focus coil (FC) module, sandwiched between
two scintillating-fibre trackers (TKU, TKD) placed in superconducting
solenoids (SSU, SSD).
Together, SSU, FC, and SSD formed the magnetic channel.
The MICE coordinate system is such that the $z$-axis is coincident
with the beam direction, the $y$-axis points vertically upwards, and the
$x$-axis completes a right-handed coordinate system.

MICE measured the passage of single particles through the apparatus which were aggregated into
a beam offline.
This paper documents the performance, during 2015-2017, of the instrumentation which was used to fully
characterise the beam and its evolution along the magnetic channel,
and quantifies the physical properties of the liquid hydrogen absorber.
The beam instrumentation consisted of three time-of-flight detectors
(TOF0, TOF1, TOF2) discussed in section~\ref{Sect:TOF}, two 
threshold Cherenkov counters (CkovA, CkovB) discussed in
section~\ref{Sect:Ckov}, a sampling calorimeter (KL) discussed in
section~\ref{Sect:KL}, a tracking calorimeter (EMR) discussed in
section~\ref{Sect:EMR}, and the scintillating-fibre trackers discussed in section~\ref{Sect:Tracking}.
The properties of the liquid hydrogen
absorber are described in section~\ref{Sect:Absorber}.

\graphicspath{{02-TOF/Figures/}}

\newcommand{\Tzero}{\ensuremath{T0}}
\newcommand{\Gauss}{\ensuremath{\text{G}}}
\newcommand{\Dt}{\ensuremath{\Delta t}}
\newcommand{\us}{\ensuremath{\mu\text{s}}}

\section{Time-of-flight detectors}
\label{Sect:TOF}

Three scintillator hodoscopes were used:
to measure the time of flight (TOF) of the particles that made up the beam;
to measure the transverse position at which the particle crossed each of the detectors;
and to provide the trigger for the experiment. 
TOF0 and TOF1~\cite{NOTE145,NOTE241,2010NIMPA.615...14B} were
placed upstream of the magnetic channel, while TOF2~\cite{NOTE286}
was located downstream of the channel, mounted in front of the KL
pre-shower detector (see figure~\ref{fig:BL}).
%The range of particle momentum delivered to the experiment was
%determined by the settings of the two dipole magnets D1 and D2.
At 240\,MeV/$c$, the difference in the TOF for a muon and a
pion between TOF0 and TOF1 was about 1.3\,ns.
The system was therefore designed to measure the TOF with a
precision of 100\,ps. 
This allowed the TOF between the first pair of TOF stations 
to be used to discriminate between pions, muons, and electrons,
contained within the beam, with near 100\% efficiency~\cite{2016JInst..11P3001A}.
In addition, by assuming a mass hypothesis for each particle, the
TOF measurement was used to infer the particle
momentum.
The TOF detectors, which operated smoothly during the running periods,
were essential for all the measurements that were
performed~\cite{Bogomilov:2012sr,Adams:2013lba,2015JInst..10P2012A,2016JInst..11P3001A,Adams:2018qhj,Bogomilov:2019kfj}.

\begin{figure}[htb]
  \begin{center}
    \includegraphics[width=7.4cm]{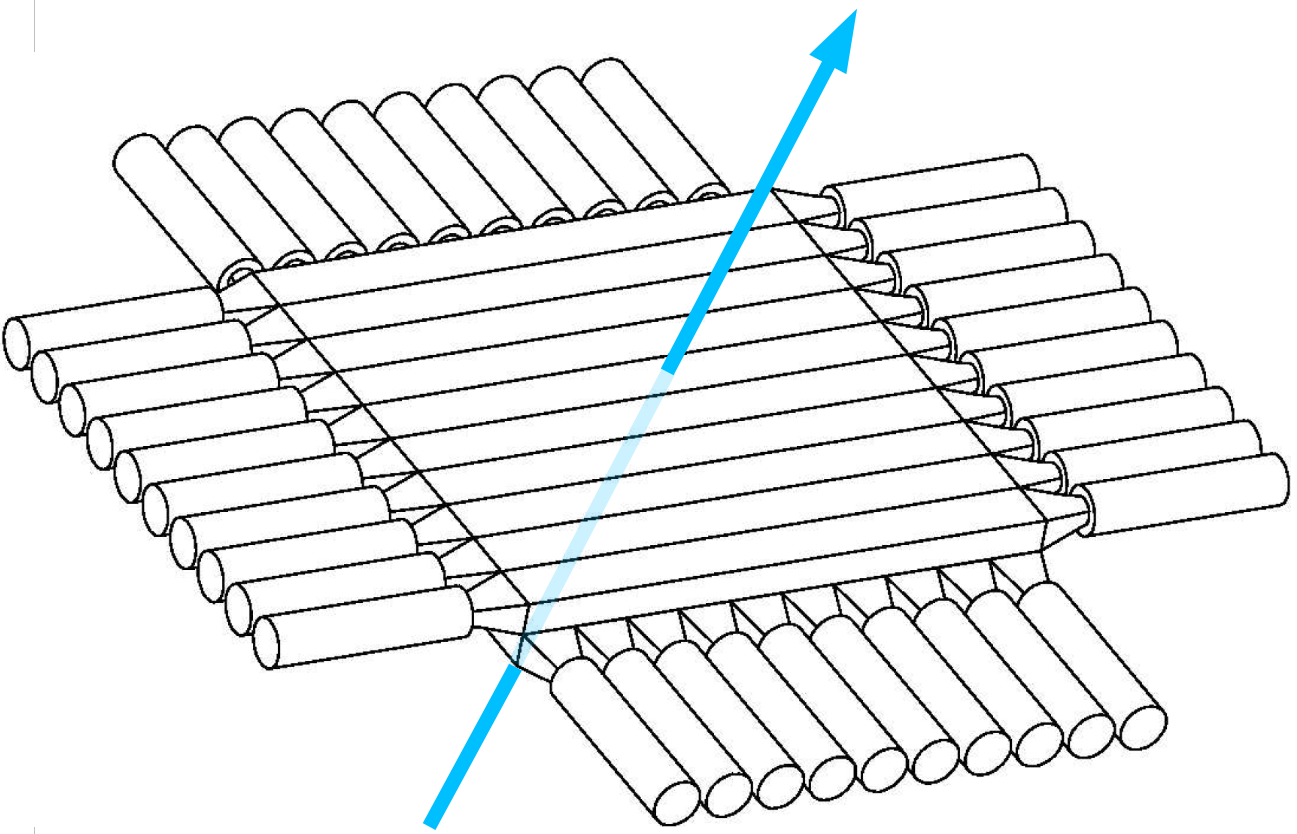}
    \includegraphics[height=2.8cm]{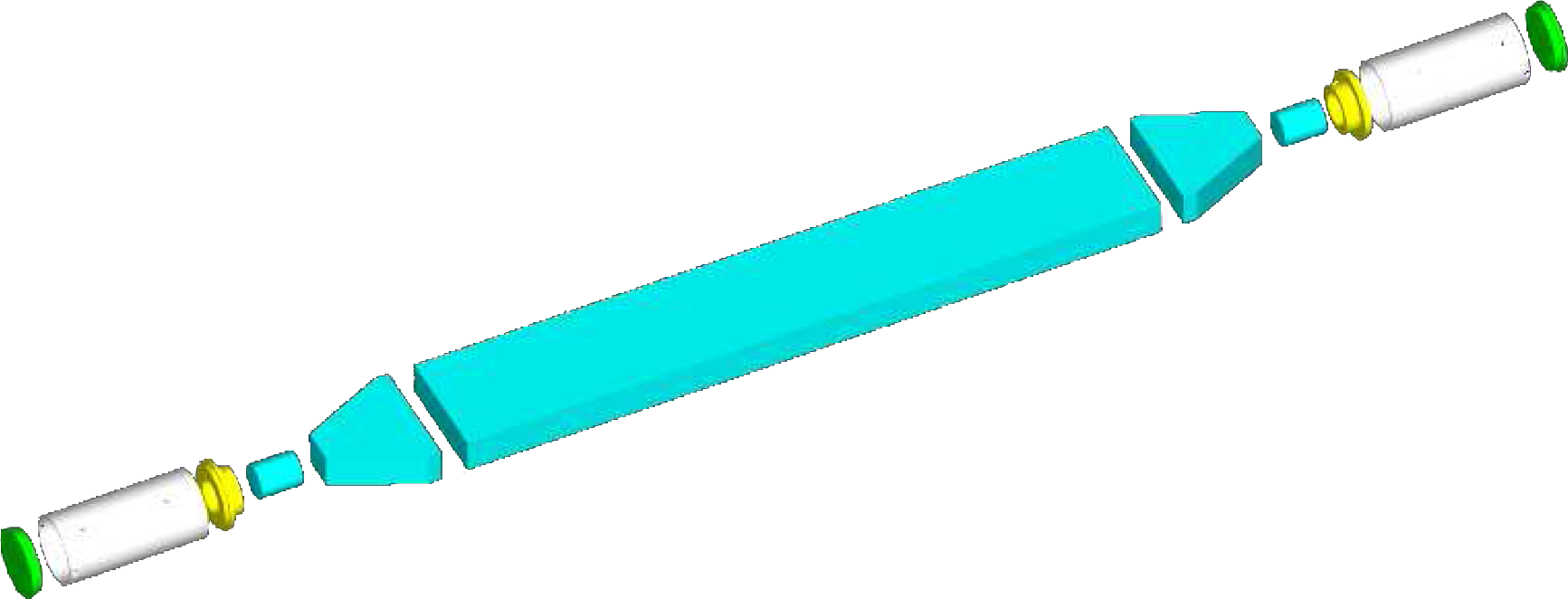}
  \end{center}
  \caption{
    The structure of the time-of-flight
    detectors~\cite{2010NIMPA.615...14B,NOTE145} showing the
    horizontal and vertical layers of slabs (left) and an exploded
    view of each slab (right). 
    The components of each slab are the central scintillator bar, two
    fishtail, clear plastic light-guides coupled to clear plastic
    matching pieces, and two PMTs.
    The beam direction is represented by the blue arrow perpendicular to the slabs.
  }
  \label{fig:tof:schematic}
\end{figure}

Each TOF station was made of two planes of 1\,inch thick scintillator
bars oriented along the $x$ and $y$ directions. 
The bars of TOF0 (TOF1, TOF2) were made of Bricon BC-404 (BC-420) plastic scintillators.
A simple fishtail light-guide was used to attach each end of each bar
to Hamamatsu R4998 fast photomultiplier tubes (PMTs).
Each PMT was enclosed in an assembly that included the voltage divider
chain and a 1\,mm thick $\mu$-metal shield.
For TOF1 and TOF2 an additional soft iron (ARMCO) local shield was
also used~\cite{Bonesini_2012,NOTE455}.
The shield was required to reduce the stray magnetic field within the
PMT to a negligible level~\cite{2010NIMPA.615...14B}.
To increase the count-rate stability, active dividers were used.
One TOF detector is illustrated in figure~\ref{fig:tof:schematic}.

The active areas of the three hodoscopes were 40$\times$40\,cm$^2$
(TOF0), 42$\times$42\,cm$^2$ (TOF1), and 60$\times$60\,cm$^2$ (TOF2).
Each of the planes in TOF0 and TOF2 had 10 slabs while those in TOF1 had 7.
A passive splitter was used to take the signal from each of the PMTs
to a LeCroy 4115 leading-edge discriminator followed by a CAEN
V1290 TDC for time measurement and to a CAEN V1724 FADC for
pulse-height measurement.
A local readout trigger was issued if the signals from each of the two
PMTs on a single slab crossed a specific threshold and overlapped.
TOF1 was used to trigger the readout of the experiment for most of the
data taking. \\

\noindent\textbf{Calibration} \\
\noindent
The intensity of the scintillation light produced when a particle
crossed the plastic scintillator rose rapidly before decaying
with a characteristic time of 1.8\,ns. 
The scintillation light travelled from the particle-crossing point to
each end of the scintillator slab.
The light's travel time depended on the distance of the particle crossing from the PMT.
The propagation speed of the light pulse along the slabs was determined to be 13.5\,cm/ns.

The local readout-trigger signal was distributed to all TDC boards and was
used as the reference time.
The time between a particle hit in a TOF slab and the time when the
trigger was generated varied with the position of the hit along the slab.
As a consequence, the reference time had an offset dependent on
the crossing position, an effect referred to as the
readout-trigger signal delay.
To compensate for this, the final time measurement in each station was
an average of the times recorded for each channel above
threshold.

Further delay was introduced by the signal-transit time of each PMT
and of the cable that led the signal to the readout electronics.
These signal-transit times were unique for each individual readout
channel and were determined by dedicated measurements.
The use of a linear, leading-edge discriminator led to a correlation
between the total charge in the pulse and the time at which the
discriminator fired.
This correlation, referred to as the time-walk, introduced a
systematic offset in the time recorded by the TDC that was dependent on the pulse
height. 

Precise determination of the TOF required a calibration procedure that
allowed channel-by-channel variations in the response of the system to
be accounted for.
The calibration procedure described in~\cite{NOTE251} accounted for
each of the effects identified above. \\

\noindent\textbf{Reconstruction} \\
\noindent
A particle crossing a TOF station passed through two orthogonal
slabs.
Signals from each PMT were corrected for time-walk, readout-trigger
signal delay, and the channel-specific delays.
The slab-crossing time was taken to be the average of the corrected
PMT times.
Two slab signals were taken to have been produced by the passage of a
particle if their slab-crossing times were within a 4\,ns window.
These two \textit{matched} slabs were used to define a pixel of area given by the
width of the slabs.
The particle-crossing time was then determined as the average of the
slab-crossing times and the approximate position of the particle crossing was
refined using the PMT signals in the two orthogonal slabs. \\

\begin{figure}[htb]
  \begin{center}
    \includegraphics[width=0.99\columnwidth]{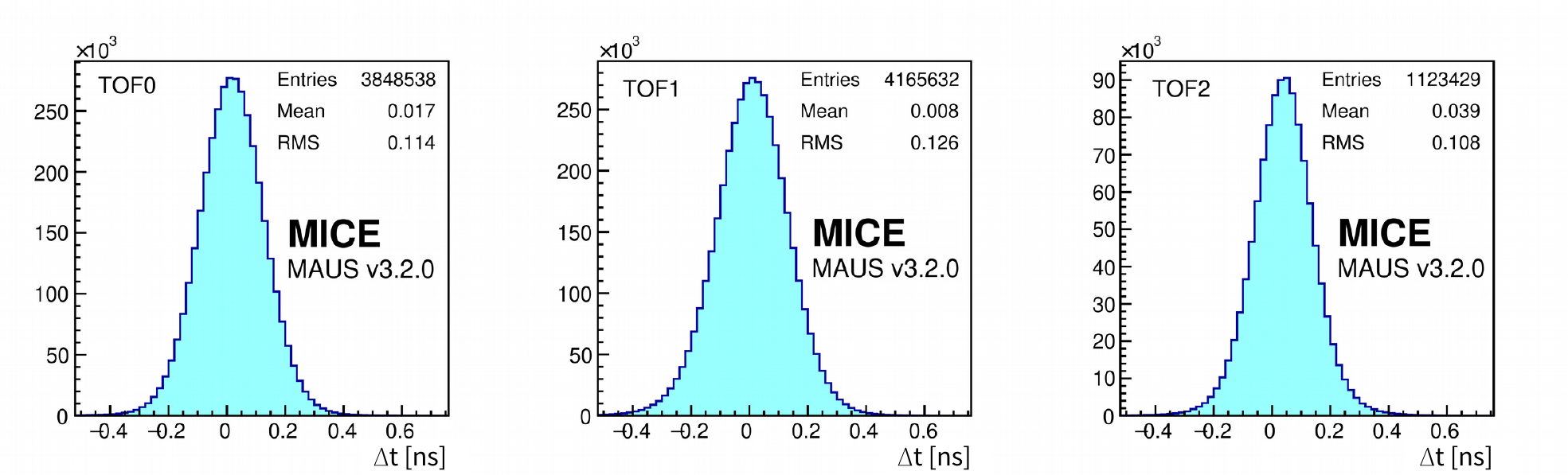}
  \end{center}
  \caption{
    Slab \Dt{} distributions.
    Total width of the distribution is due to the resolution of the
    individual channels and due to the offsets in their \Dt{}
    distributions.
  }
  \label{fig:SlabDtAll}
\end{figure}

\noindent\textbf{Performance} \\
\noindent
The difference, $\Delta t$, between the slab-crossing times for \textit{matched} slabs
was used to determine the intrinsic time resolution, $\sigma_t$ of the
TOF system.
The $\Delta t$ resolution, $\sigma_{\Delta t}$, is given by $\sigma_{\Delta t}=2\sigma_t$,
assuming that the intrinsic resolution is the same in each of the planes that make up a particular TOF station.
Figure~\ref{fig:SlabDtAll} shows the distributions of $\Delta t$ for
TOF0, TOF1, and TOF2 for a representative set of data taken in 2017.
The RMS width of the distributions are 114\,ps, 126\,ps, and 108\,ps
for TOF0, TOF1, and TOF2 respectively.
The distributions are similar, and the RMS of each distribution is
consistent with the measured intrinsic resolution of approximately
60\,ps~\cite{2010NIMPA.615...14B}.

\begin{figure}
  \begin{center}
    \includegraphics[width=0.75\columnwidth]{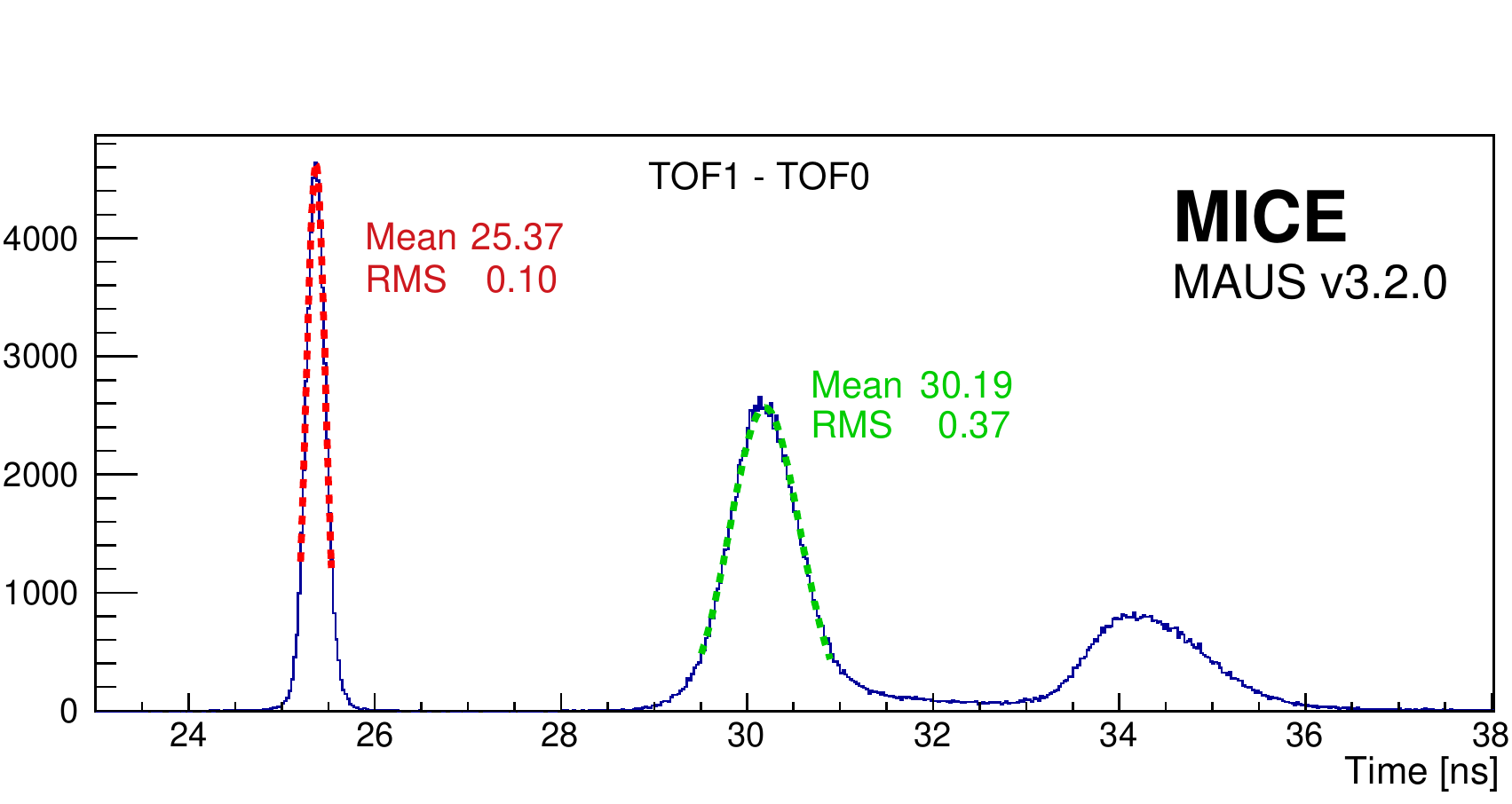}
  \end{center}
  \caption{
    Time of flight between TOF0 and TOF1 after all corrections have
    been applied.
    The electron (left-most peak, shown in red), the muon (central peak, shown in green), and the pion (right-most peak, shown in blue) peaks are clearly separated.
    } 
  \label{fig:TOF_peaks}
\end{figure}

Figure~\ref{fig:TOF_peaks} shows an example distribution of the
measured TOF between TOF0 and TOF1.
The TOF peaks characteristic of electrons, muons, and pions are
clearly separated.
The width of the electron peak is approximately~0.10\,ns,
consistent with the spread calculated from a naive quadrature addition
of the timing resolution of the individual TOF stations.

\graphicspath{{03-Ckov/Figures/}}

\section{Cherenkov Detectors}
\label{Sect:Ckov}

\begin{figure}[htb]
  \begin{center}
    \includegraphics[width=0.65\columnwidth]{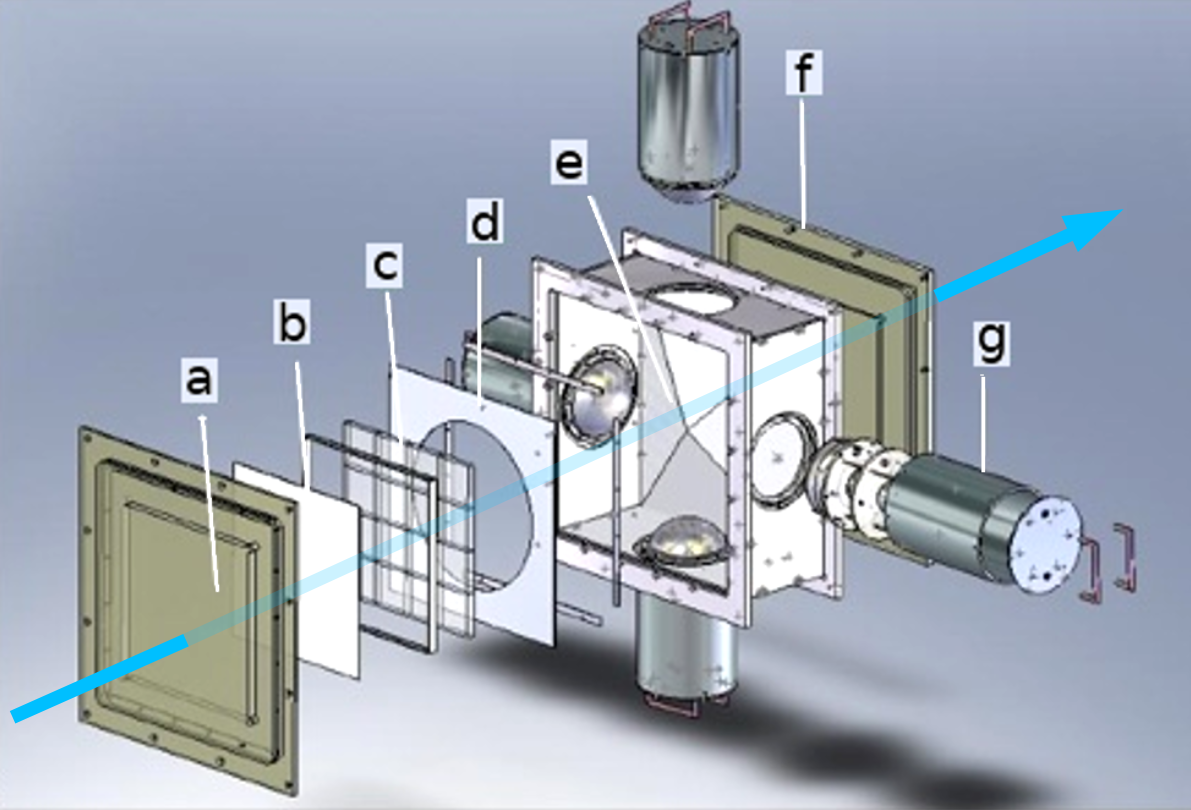}
  \end{center}
  \caption{
    MICE aerogel Cherenkov counter: a)~entrance window,
    b)~mirror, c)~aerogel mosaic, d)~acetate window, e)~GORE DRP reflector
    panel, f)~exit window and g)~eight-inch PMT in iron shield.
    The beam direction is represented by the blue arrow traversing the detector.
  } 
  \label{fig:ckov1}
\end{figure}

The threshold Cherenkov counters were designed to distinguish
muons from pions at particle 
\linebreak[4] 
momenta~$\gsim 200$\,MeV/$c$, where
the precision of the time-of-flight measurement was not sufficient for
conclusive identification.
Two high-density silica aerogel Cherenkov detectors with refractive
indices $n$=1.07 (CkovA) and $n$=1.12 (CkovB) were used~\cite{Cremaldi:2009zj}.
The structure of the detectors is shown in figure~\ref{fig:ckov1}.
Light was collected in each counter by four eight-inch, UV-enhanced
PMTs and recorded using CAEN V1731 FADCs~\cite{NOTE473}.
The two detectors were placed directly one after the other in the
beamline and located just after TOF0.

The refractive indices of CkovA and CkovB result in detection
thresholds for muons of approximately 280\,MeV/$c$ and 210\,MeV/$c$ respectively.
For pions, the thresholds are approximately 367\,MeV/$c$ (CkovA) and
276\,MeV/$c$ (CkovB).
MICE was designed to operate using beams with a central momentum
between 140\,MeV/$c$ and 240\,MeV/$c$.
%The thresholds of CkovA and CkovB were chosen to match this range
%since the TOF system was able to identify muons with high efficiency
%for momenta below 210\,MeV/c.
%For momentum greater than 210\,MeV/c and less than 276\,MeV/c, above
%the maximum beam momentum, muons will produce a signal in CkovB while
%pions will produce no signal in either detector.
The Cherenkov counters' thresholds were chosen to provide muon identification for beams of 210\,MeV/$c$ and above, while the TOFs provide muon identification for beam below 210\,MeV/$c$.
Unambiguous identification of particle species using the Cherenkovs
exploited the momentum measurement provided by the trackers. \\

\begin{figure}[htb]
  \begin{center}
    \includegraphics[width=0.99\columnwidth]{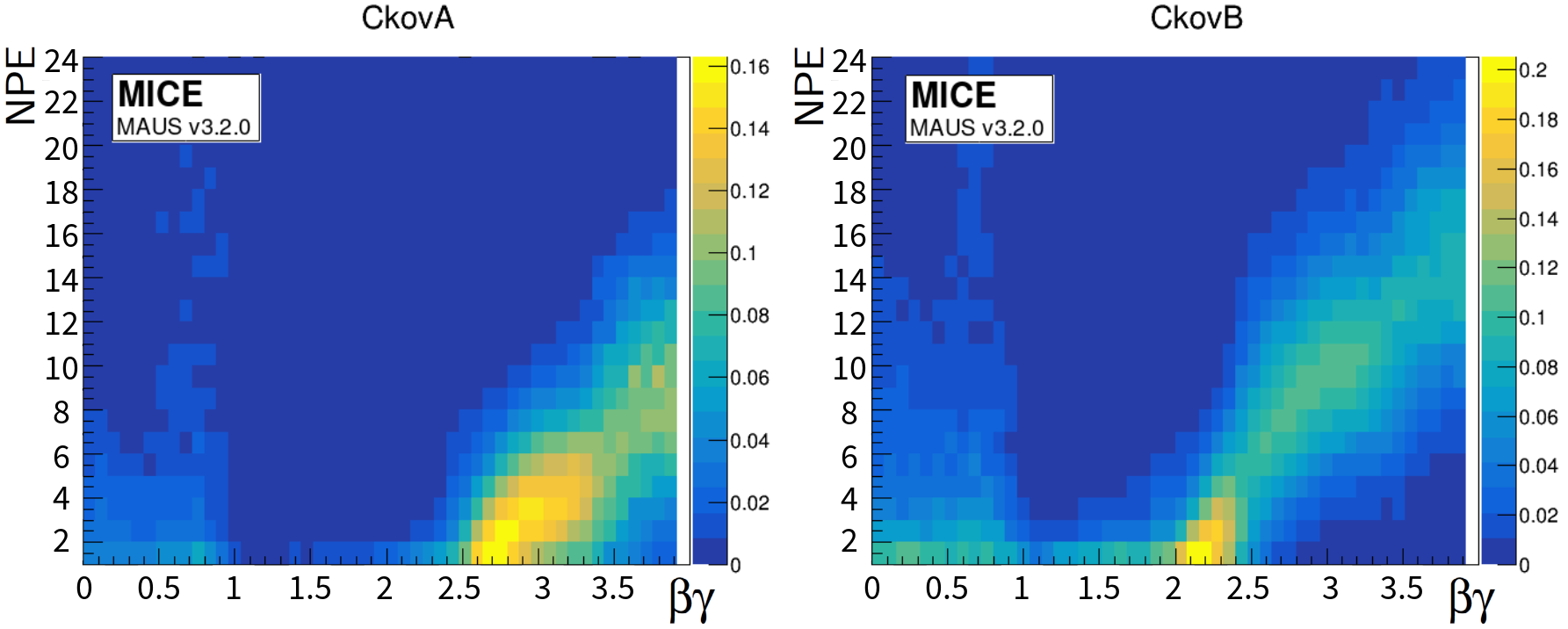}
    \caption{Photoelectron yields versus $\beta\gamma$ in CkovA and CkovB,
    where $\beta c$ is the particle velocity and $\gamma=(1-\beta^2)^{-\frac{1}{2}}$.
    }
    \label{fig:ckov_betagamma}
  \end{center}
\end{figure}

\noindent\textbf{Performance} \\
\noindent
The performance of the detectors was determined using beams for which
the momentum range was broad enough to observe the turn-on points and to
allow the asymptotic light yields (as the particle velocity divided by the speed of light, $\beta$, approaches 1) to be
obtained from fits to the data.
The normalised photo-electron yields observed in CkovA and CkovB are
plotted as a function of $\beta\gamma$ (where $\gamma=(1-\beta^2)^{-\frac{1}{2}}$) in
figure~\ref{fig:ckov_betagamma}.
The pedestal in the photo-tube response arising from background
photons has been subtracted.
The approximate turn-on points for CkovA and CkovB were found at
$\beta\gamma \approx 2.6$ and $\approx 2.1$ respectively,
corresponding to refractive indices of $n \approx 1.07$ and $\approx 1.11$ which are in broad agreement with the
properties of the aerogel radiators. 

\graphicspath{ {04-KL/Figures/} }

\section{KLOE-Light Calorimeter}
\label{Sect:KL}

\begin{figure}
  \begin{center}
    \includegraphics[width=0.85\columnwidth]{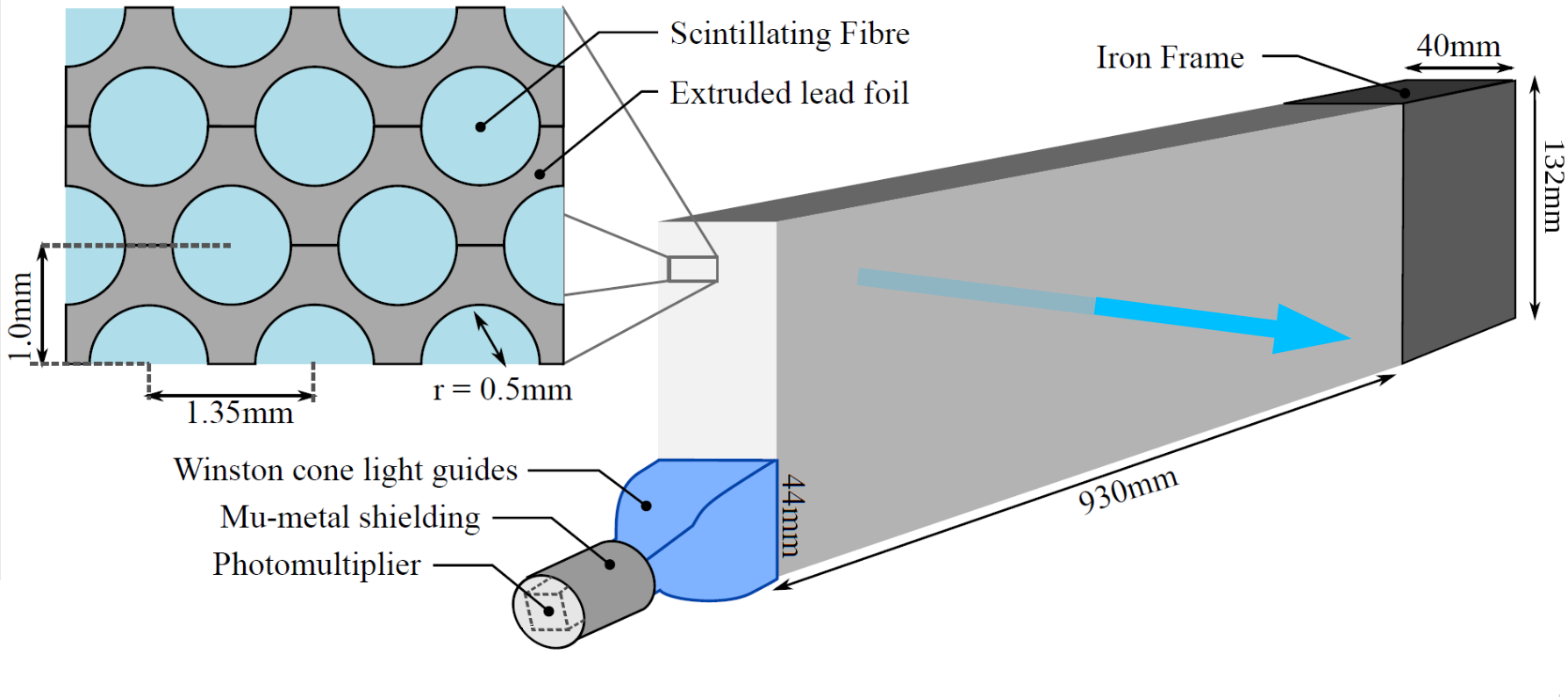}
    \caption{Single slab design of MICE KLOE-Light Calorimeter~\cite{Overton:2014tka}; only one of the six PMT assemblies is shown.
    The beam direction is represented by the blue arrow traversing the slab.
    }
    \label{fig:KL2}
  \end{center}
\end{figure}

The KLOE-Light (KL) pre-shower sampling calorimeter was composed of
extruded lead foils in which scintillating fibres were placed.
At normal incidence the thickness of the detector was 2.5 radiation
lengths.
The detector provided energy deposition and timing information and was
used to distinguish muons from decay
electrons~\cite{2016JInst..11P3001A}.
The KL consisted of a series of layers of 1\,mm diameter BICRON BCF-12
scintillating fibres embedded in an appropriately shaped lead sheets
(see figure~\ref{fig:KL2}).
Each fibre was separated by 1.35\,mm from its neighbours within a
layer and the distance between the centres of the fibres in adjacent
layers was 0.98\,mm.
One layer was shifted by half the fibre pitch with respect to the next.
The volume ratio of scintillator to lead was approximately 2:1,
``lighter'' than the ratio of 1:1 used in the similar calorimeter of the KLOE
experiment~\cite{Ambrosino:2009zza}. 
Lead/scintillator layers were stacked into slabs, 132\,mm in depth.
A total of 7 slabs formed the whole detector, which had an active
volume of 93\,cm$\times$93\,cm$\times$4\,cm.
Scintillation light was guided from each slab into a total of six PMTs
(three at each end).
Iron shields were fitted to each photomultiplier to mitigate the effect of stray magnetic fields.
The signal from each PMT was sent to a shaping amplifier module
that stretched the signal in time to match the sampling rate
of the CAEN 1724 FADCs. \\

\noindent\textbf{Performance} \\
\noindent
%The response of the KL to muons, pions, and electrons is shown for various beam momentum settings in~figure~\ref{fig:KL3}.
To study the response of the KL, the particle momentum was determined from the
measured time-of-flight between TOF0 and TOF1.
To compensate for the effect of attenuation the performance was
evaluated in terms of the ``ADC product'' given by:
\begin{equation}
  \text{ADC}_{\text{prod}} = \frac{2 \times
    \text{ADC}_{\text{left}} \times \text{ADC}_{\text{right}}}{
    (\text{ADC}_{\text{left}} + \text{ADC}_{\text{right}})}\,;
\end{equation}
where ADC$_{\text{left}}$ and ADC$_{\text{right}}$ are the signals
from the two ends of a slab and the factor of 2 is present for
normalisation.
Data was taken with no field in the spectrometer solenoids or the
focus coil at beam-momentum settings chosen to span the range of
momenta used during MICE running.
The resulting momentum distributions were centred at 140, 170,
200, 240, and 300\,MeV/$c$.
The response of the KL to muons and pions was observed to increase with
beam momentum.
%\begin{figure}
%  \begin{center}
%    \includegraphics[width=0.45\columnwidth]{./04-KL/Figures/muon-edited.png}
%    \includegraphics[width=0.45\columnwidth]{./04-KL/Figures/pion-edited.png}
%    \includegraphics[width=0.45\columnwidth]{./04-KL/Figures/electron-edited.png}
%  \end{center}
%  \caption{
%    KL response to muons (top left), pions (top right) and electrons
%    (bottom) for all available momenta.
%    The charge deposited by particles in KL in arbitrary units is
%    shown.
%    140\,MeV/$c$ pions are irrelevant at the KL position.
%    All histograms are normalised to unity.
%  }
%  \label{fig:KL3}
%\end{figure}
  
\begin{figure}
  \begin{center}
    \includegraphics[width=0.49\columnwidth]{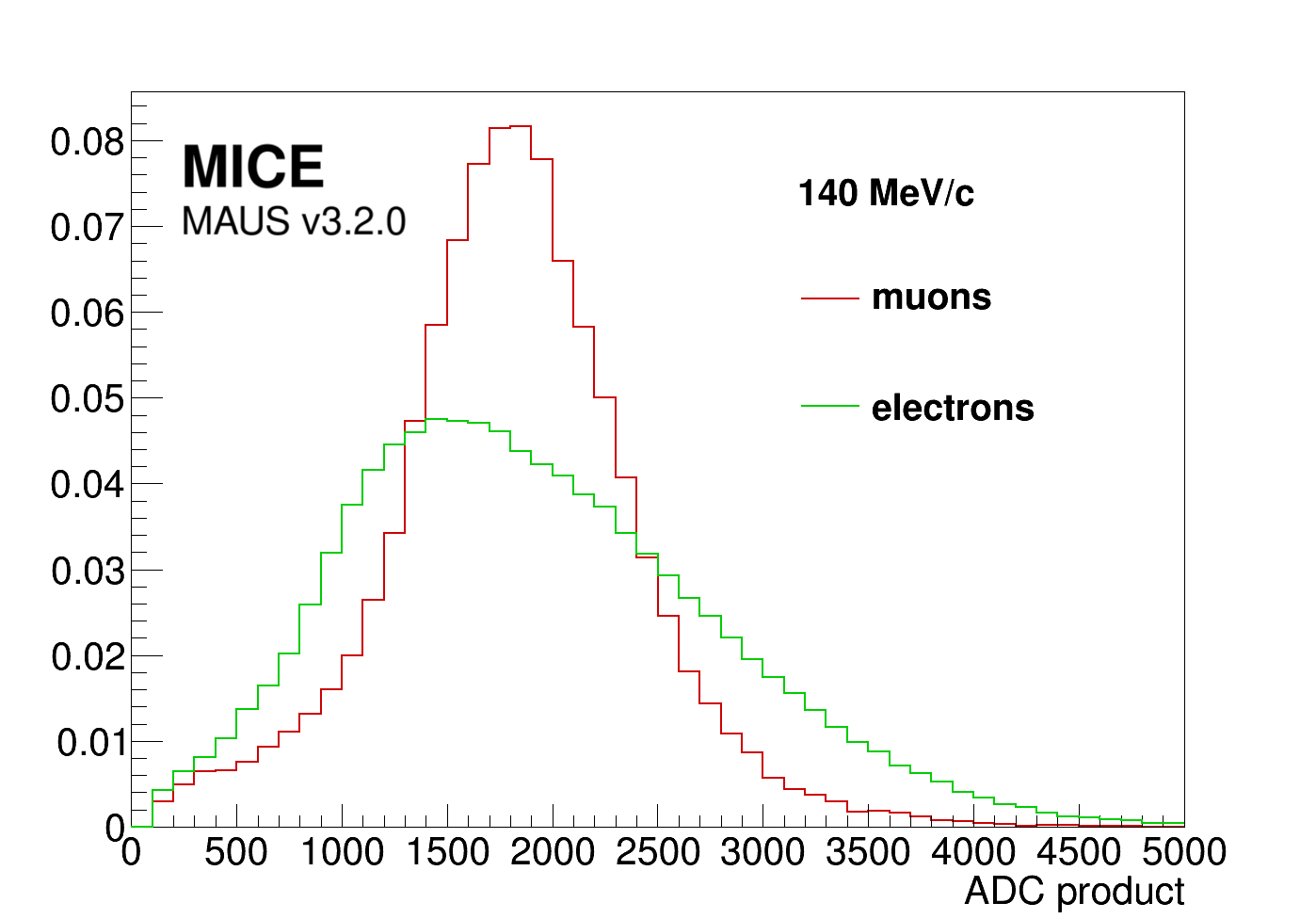}
    \includegraphics[width=0.49\columnwidth]{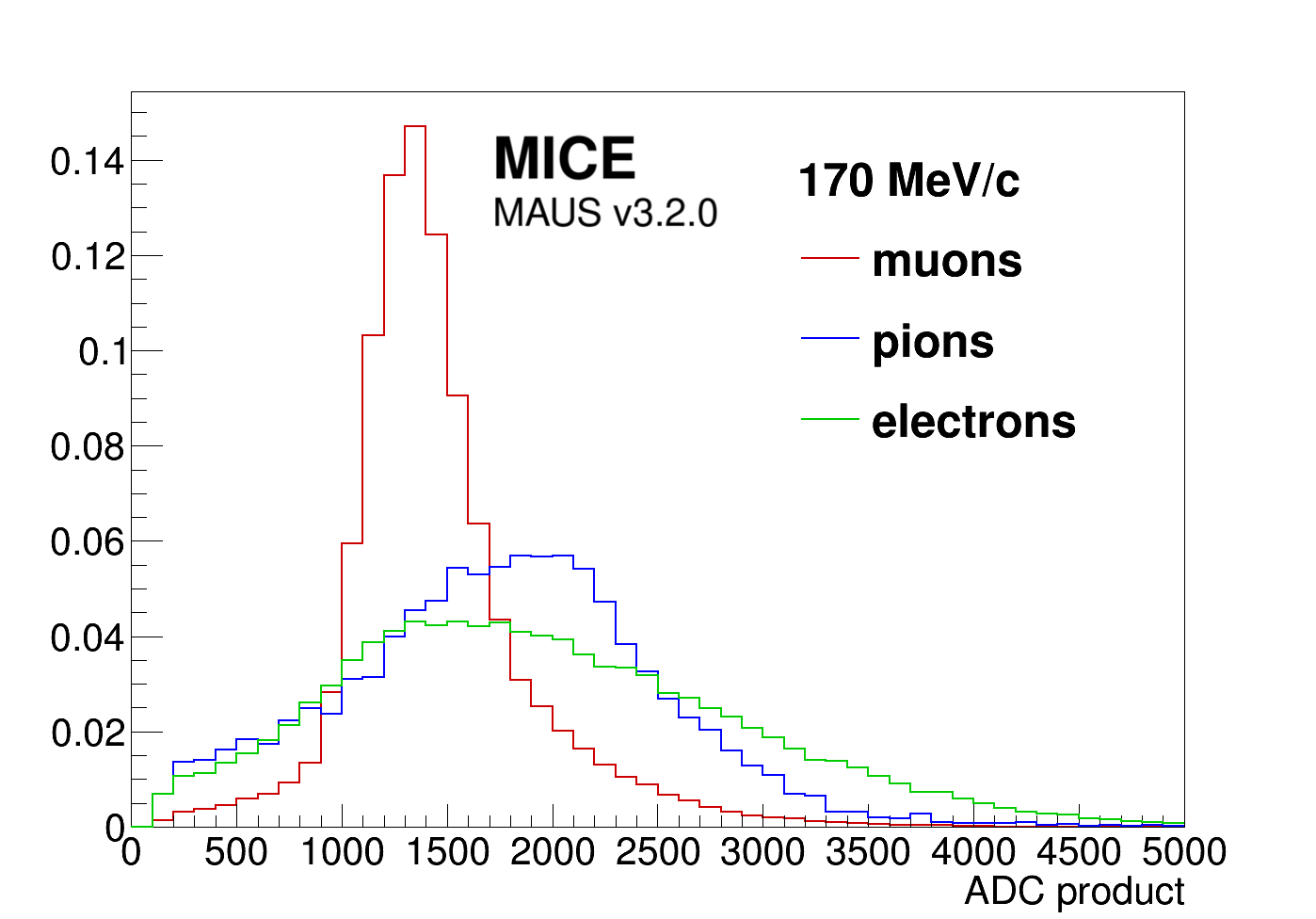} 
    \includegraphics[width=0.49\columnwidth]{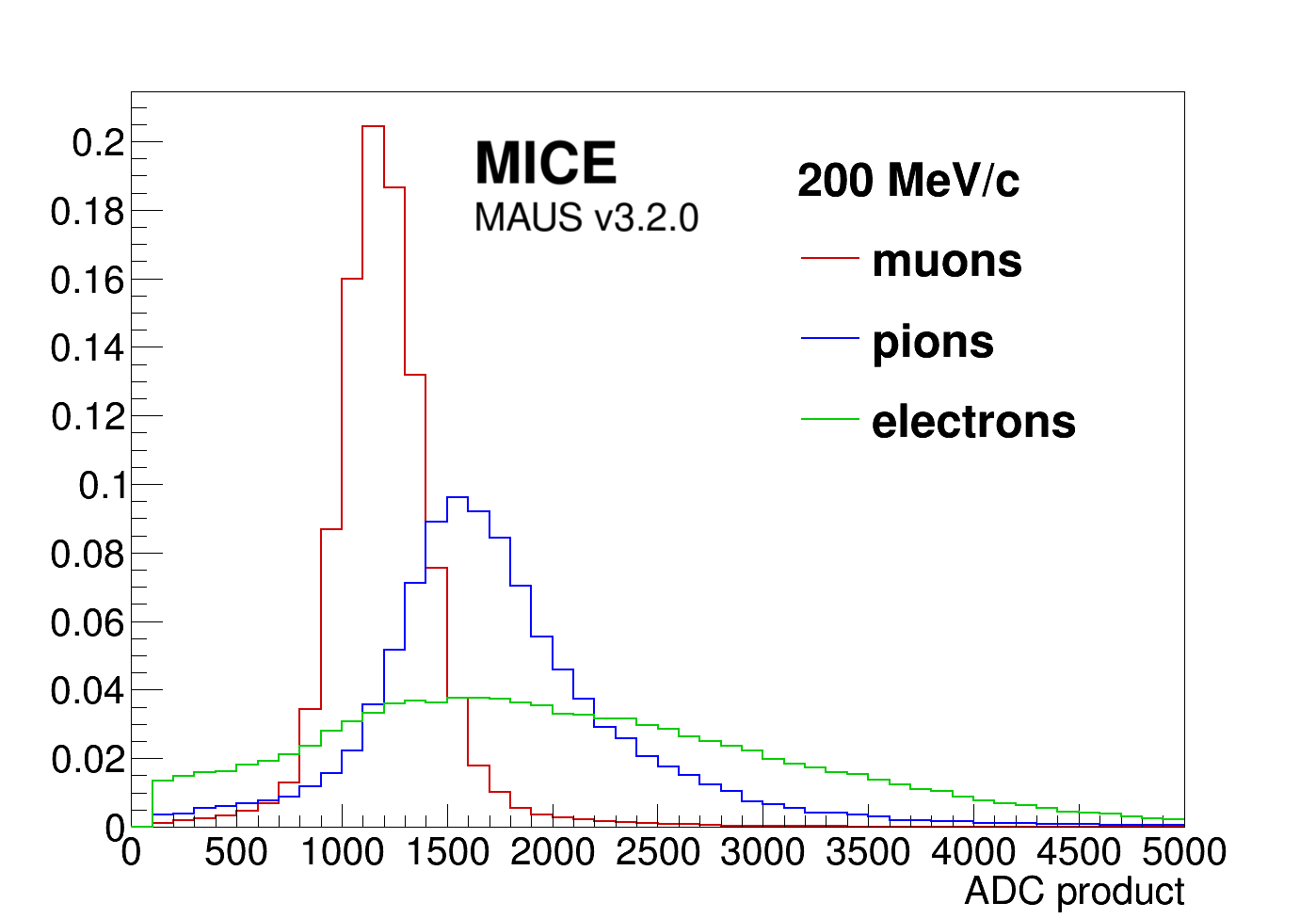}
    \includegraphics[width=0.49\columnwidth]{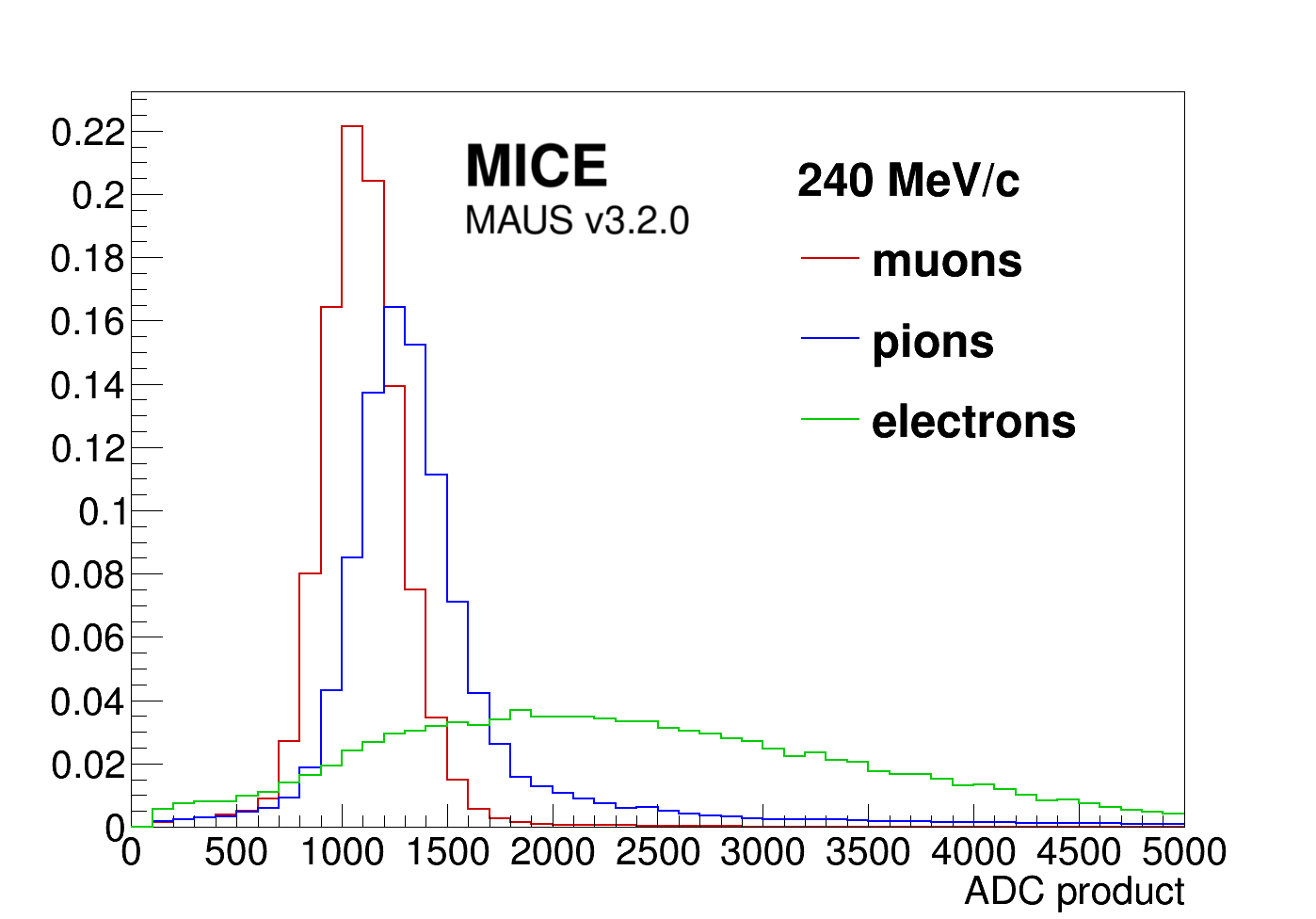}  	
    \includegraphics[width=0.49\columnwidth]{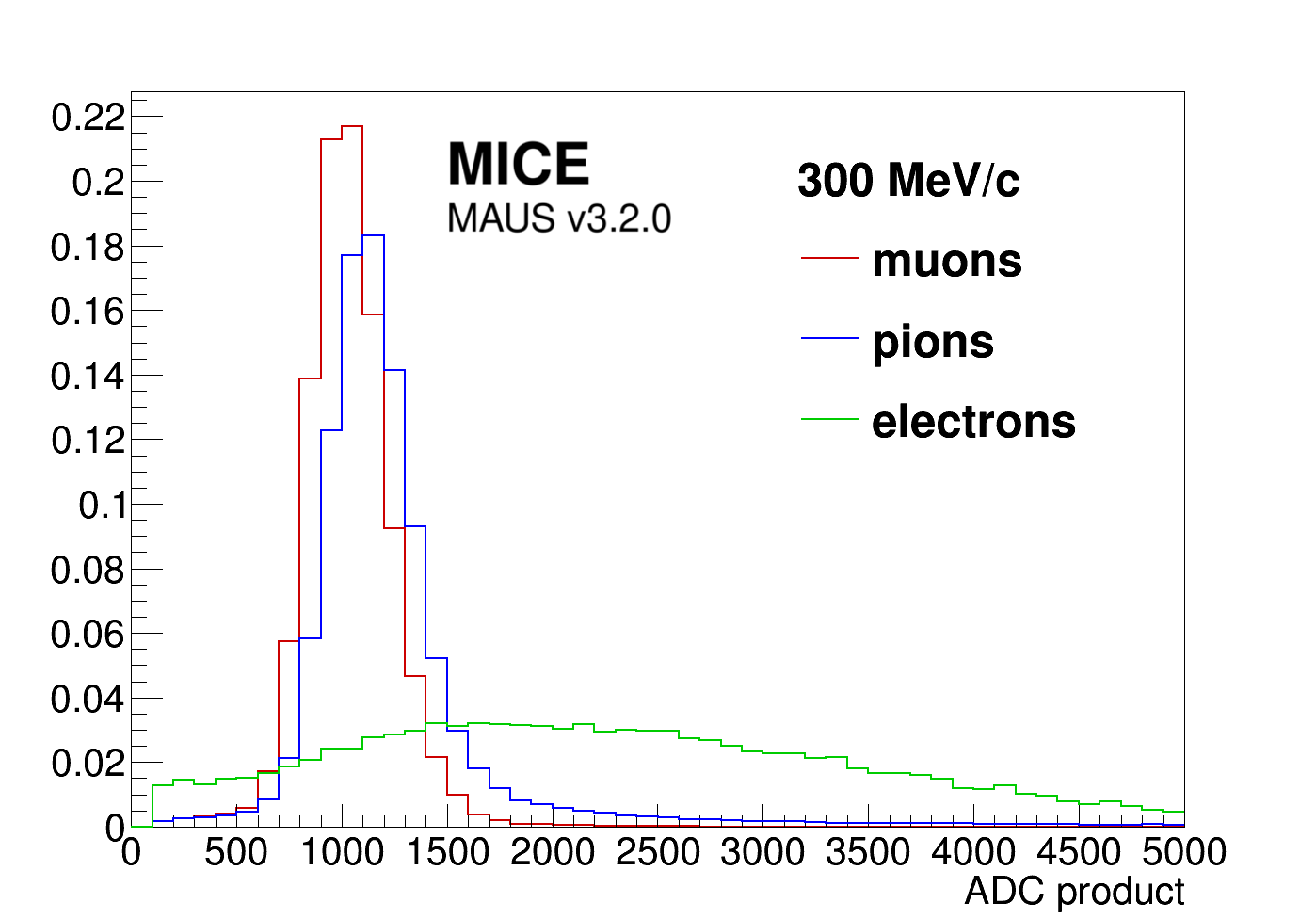}
  \end{center}
  \caption{
    Comparison of ADC products of muons (red), pions (blue) and electrons (green) traversing the KL, at 140
    MeV/$c$ (top left), 170 MeV/$c$ (top right), 200 MeV/$c$ (middle
    left), 240 MeV/$c$ (middle right) and 300 MeV/$c$ (bottom).
  }
  \label{fig:KL4}
\end{figure}

Figure~\ref{fig:KL4} presents a comparison of the response to muons,
pions and electrons for various beam momentum settings.
At high momentum, for example 300\,MeV/$c$, the ADC~product
distributions for muons and pions are similar.
At lower momentum the distributions become increasingly dissimilar,
the pions having a broader distribution arising from hadronic
interactions.
The difference between the detector's response to pions and muons has
been exploited to determine the pion contamination in the muon beams
used for the MICE cooling measurements~\cite{2016JInst..11P3001A}.  
\begin{figure}
  \begin{center}
    \includegraphics[width=0.49\columnwidth]{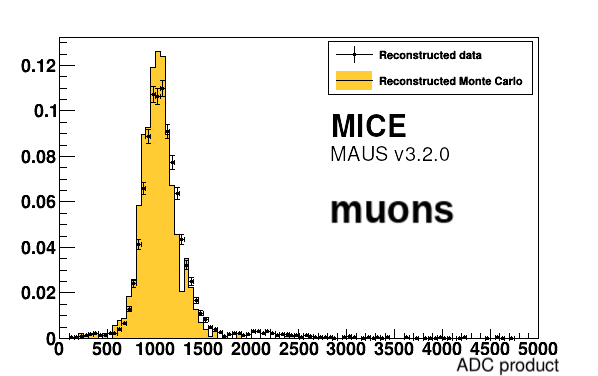}
    \includegraphics[width=0.49\columnwidth]{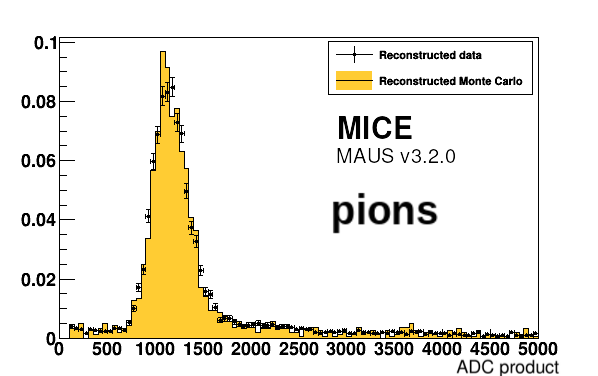}
  \end{center}
  \caption{
    Comparison between data and Monte Carlo simulation of KL response
    to muons (left) and pions (right) at 300 MeV/$c$.
  } 
  \label{fig:KL_mc_vs_data}
\end{figure}

The ADC~product distribution measured using a 300\,MeV/$c$ beam is
compared to the MAUS~\cite{MAUS} simulation of the detector response in
figure~\ref{fig:KL_mc_vs_data}.
The simulation takes into account the light production distribution of the
scintillating fibres, and the response of the PMTs for which the gain was
approximately $2 \times 10^6$. 
The data is well described by the simulation.
 
\graphicspath{{05-EMR/Figures/}}

\section{Electron Muon Ranger}
\label{Sect:EMR}

\begin{figure}[htb!]
  \begin{center}
    \includegraphics[width=0.465\columnwidth]{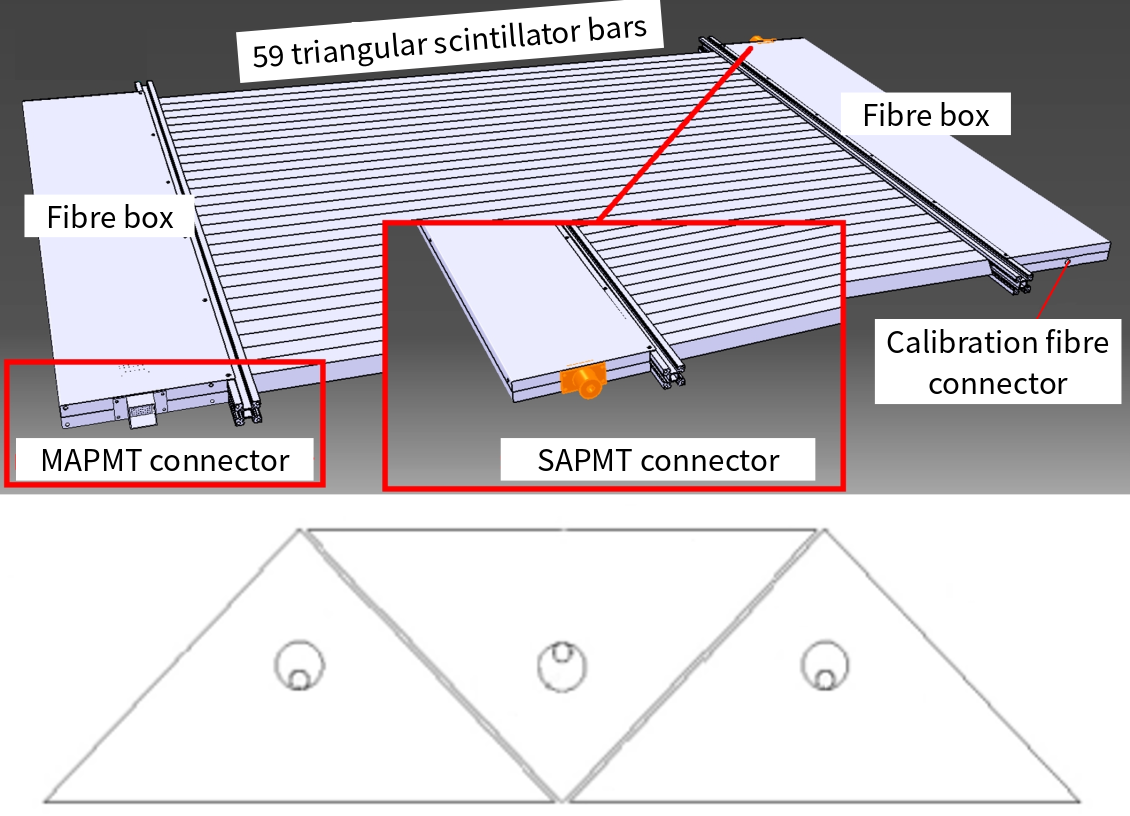}
    \hfill
    \includegraphics[width=0.515\columnwidth]{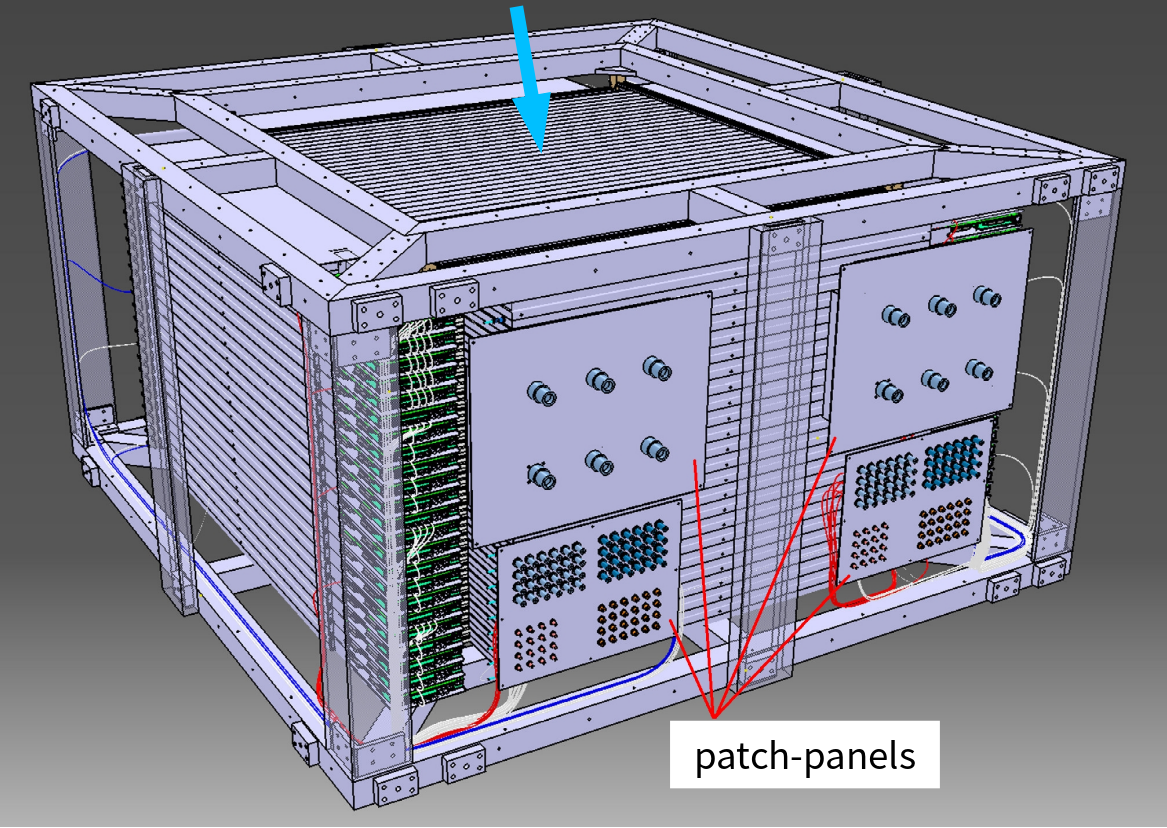}
  \end{center}
  \caption{
    Drawing of one EMR plane (top left), cross section of the
    arrangement of 3 bars and their wavelength shifting fibres (bottom
    left) and drawing of the full detector and its supporting
    structure from a top perspective (right).
    The beam direction is represented by the blue arrow perpendicular to the detector.
  }
  \label{fig:EMR}
\end{figure}

The EMR was a fully-active scintillator
detector~\cite{2016JInst..11T10007} with a granularity that allowed track reconstruction.
The EMR consisted of extruded triangular scintillator bars arranged in
planes.
Each plane contained 59 bars and covered an area of 1.27\,m$^2$.
Figure~\ref{fig:EMR} shows the bar cross section and the arrangement of the
bars in a plane.
Triangular bars were chosen so that tracks moving parallel to the
detector axis could not travel along the gaps between bars. 
Successive planes were mounted perpendicularly, so
that hits in neighbouring planes defined a position.
A single ``X--Y module'' was a pair of orthogonal planes.
The scintillation light was collected using a wavelength shifting
(WLS) fibre glued inside each bar.
At each end, the WLS fibre was coupled to clear fibres that
transported the light to a PMT.
All the WLS fibres from one edge of a plane were read out using one
single-anode PMT (SAPMT) so that an integrated charge measurement could be
used to determine the energy deposited in the plane.
The signals from the fibres emerging from the other edge of the plane
were recorded individually using multi-anode PMTs (MAPMTs). 
The full detector was made up of 24 X--Y modules giving a total active 
volume of approximately~1\,m$^3$.

Measurements of the performance of the completed detector demonstrated
an efficiency per plane
of~$99.73\pm0.02$\%~\cite{2016JInst..11T10007,Drielsma:2017doj}.
The level of crosstalk was within acceptable values for the type of
MAPMT used, with an average of $0.20\pm0.03$\%
between adjacent channels and a mean amplitude
equivalent to $4.5\pm0.1$\% of the primary signal.
Only four dead bars were present.

The primary purpose of the EMR was to distinguish between a muon that crossed  the entire magnetic channel and those which decayed in flight producing an electron.
Muons and electrons exhibited distinct behaviours in the detector.
A muon produced a single straight track before either stopping or
exiting the scintillating volume.
Electrons showered in the lead of the KL and created a broad cascade
of secondary particles.
Two main geometric variables, the ``plane density" and the ``shower spread",
were used to differentiate them.
The detector was capable of identifying electrons with an efficiency
of 98.6\%, providing a purity for the MICE beam that exceeds
99.8\%.
The EMR also proved to be a powerful tool for the reconstruction of
muon momenta in the range
100--280\,MeV/$c$~\cite{2015JInst..10P2012A}.  \\

\noindent\textbf{Performance} \\
\noindent
A full description of the detector and the reconstruction algorithms
used may be found in reference~\cite{2015JInst..10P2012A}.
Here the performance of the EMR detector over the course of the
experiment is summarised.

To measure the performance of the EMR the MICE beamline was set to
deliver a nominal momentum of 400\,MeV/$c$. This maximised the muon
transmission to the EMR and its range in the detector.
In this configuration the beamline produced pions and muons in
comparable quantities, as well as a smaller number of electrons.
Time-of-flight between TOF1 and TOF2 was used to identify particle
species and only particles compatible with the muon hypothesis were
included in the analysis.
Particles entering the muon sample had a momentum larger than
350\,MeV/$c$ at the upstream surface of TOF2 and were expected to
cross both TOF2 and the KL and penetrate the EMR.
$99.62\pm0.03\%$ of the particles entering TOF2 were observed to produce
hits in the EMR.
The small inefficiency may be attributed to pions in the muon sample
that experienced hadronic interactions in the KL.
If hits were produced in the detector, an $(x,y)$ pair, defining a
space point, was reconstructed $98.56\pm0.06\%$ of the time.

To evaluate the efficiency of the scintillator planes, only the muons
that traversed the entire detector were used.
Muons were selected which produced a hit in the most downstream plane.
For these events a hit was expected in at least one bar in each plane
on its path.
The mode of the hit-multiplicity distribution per plane was one,
in $3.26\pm0.02\%$ of cases a plane traversed by a muon did not
produce a signal in the MAPMT, and the probability that the track was
not observed in the SAPMT was $1.88\pm0.01\%$. \\

\noindent\textbf{Electron rejection} \\
\noindent
A broad range of beamline momentum settings was used to characterise
the electron-rejection efficiency.
Particle species were characterised upstream of the EMR using the
time-of-flight between TOF1 and TOF2.
For each momentum setting, a fit was carried out to determine
the position of the muon and electron time-of-flight peaks and events were
selected accordingly to form muon and electron-template samples.
%For each momentum setting, a Gaussian fit was carried out to determine
%the position of the muon and electron time-of-flight peaks.
%Events which fell within {\color{red} XX standard deviations} of the
%central value of the muon time-of-flight peak were were accepted into
%a muon-template sample while events which fell within {\color{red} XX
%standard deviations} of the electron peak formed the electron-template
%sample. 
Particles with a time-of-flight larger than the upper limit of the
muon sample were either pions or slow muons and were rejected.

To distinguish the muon tracks from the electron-induced showers,
two particle-identification variables were defined based on the
distinct characteristics of the two particle species.
The first is the plane density, $\rho_p$:
\begin{equation}
  \rho_p = \frac{N_p}{Z_p+1},
\end{equation}
where $N_p$ is the number of planes hit and $Z_p$ the number of the
most downstream plane~\cite{2015JInst..10P2012A}.
A muon deposits energy in every plane it crosses until it stops,
producing a plane density close to one.
An electron shower contains photons that may produce hits deep inside
the fiducial volume without leaving a trace on their path, reducing
the plane density.
The second variable is the normalised $\hat{\chi}^2$ of the fitted
straight track given by
\begin{equation}
  \hat{\chi}^2=\frac{1}{N-4}\sum_{i=1}^{N}\frac{\text{res}_{x,i}^2+\text{res}_{y,i}^2}{\sigma_x^2+\sigma_y^2};
\end{equation}
where $N$ is the number of space points (one per bar hit),
$\text{res}_{q,i}$ the residual of the space point with respect to the
track in the $qz$ projection and $\sigma_q$ the uncertainty on the
space point in the $qz$ projection, $q=x,\,y$~\cite{Drielsma:thesis}.
This quantity represents the transverse spread of the hits produced by
the particle in the EMR.
A muon produced a single track giving
$\hat{\chi}^2$ close to one, while an electron shower produced a larger value.
The two discriminating variables can be combined to form a statistical
test on the particle hypothesis. 
Dense and narrow events will be tagged as muons while non-continuous
and wide showers will not.  
The quality of this statistical test was characterised in terms of
the fraction of the muon sample that is rejected, $\alpha$, and the fraction of the electron sample that is selected, $\beta$.

%The downstream tracker allows the reconstruction of particle momentum
%upstream of the EMR. 
%To assess the influence of momentum on contamination and loss, their
%values were calculated in 10\,MeV/$c$ momentum bins in the range
%100--300\,MeV/$c$.

\begin{figure}
  \begin{center}
    \includegraphics[width=0.68\columnwidth]{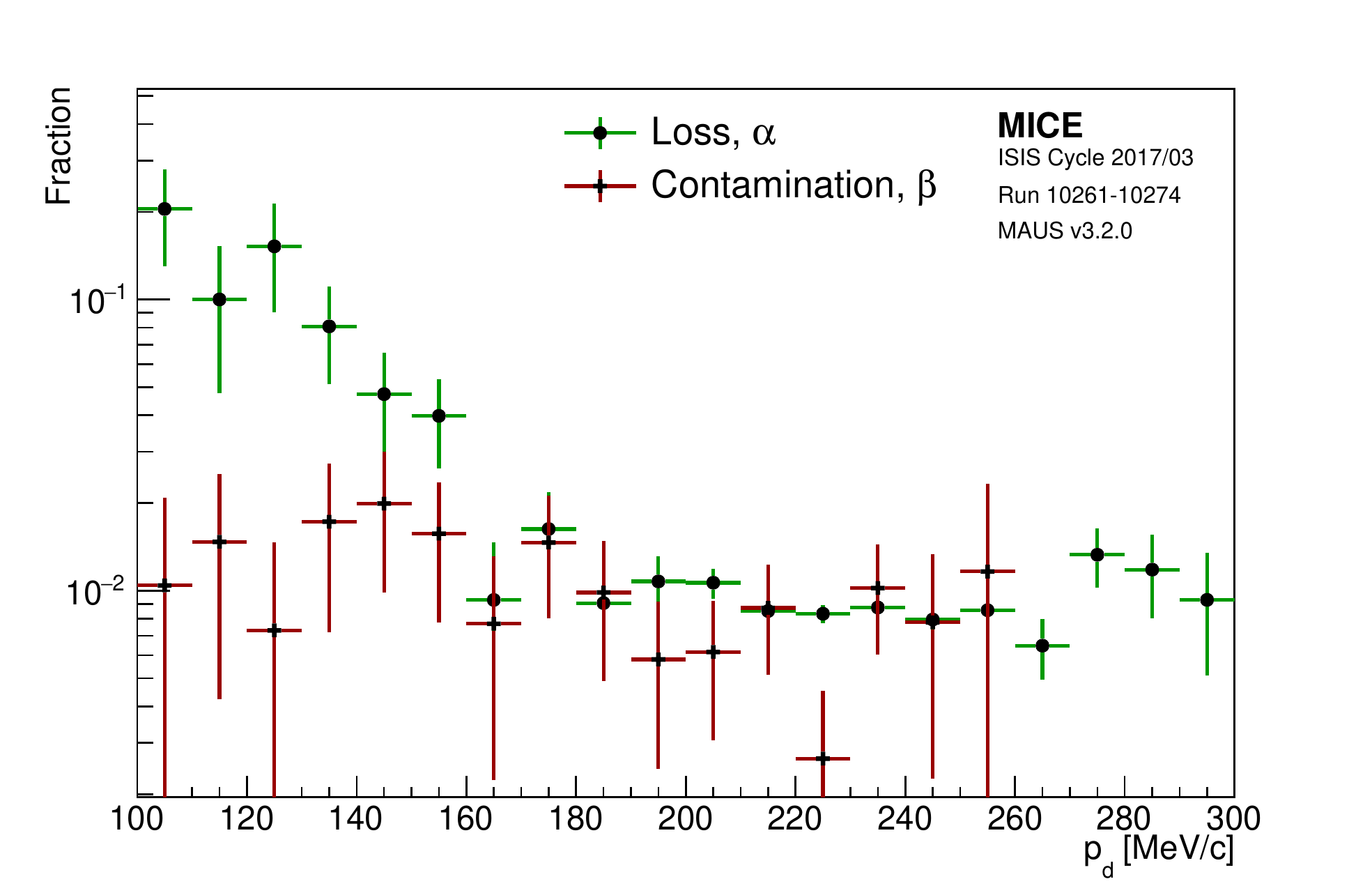}
  \end{center}
  \caption{
    Percentage of electron contamination, $\beta$, and muon loss,
    $\alpha$, for different ranges of momentum measured in the
    downstream tracker, $p_d$.
    The error bars are based on the statistical uncertainty in a bin,
    and the bin width set by the resolution of the measurement.
  }
  \label{fig:emr_pid_mom}
\end{figure}

The momentum of the particles was measured by the downstream tracker 
and this information used to determine the momentum dependence of the 
contamination and loss in the range 100--300\,MeV/$c$.
Figure~\ref{fig:emr_pid_mom} shows the loss, $\alpha$, and the
contamination, $\beta$, as a function of the momentum measured in
TKD.
$\alpha$ increases towards low muon momentum.
This is due both to an increase in the decay probability between TOF2
and the EMR and a decrease in the number of muons that cross the KL to
reach the EMR. 

\section{Tracking}
\label{Sect:Tracking}

The MICE instrumentation allowed individual particles to be tracked from TOF0 to
the EMR, a distance of more than 15\,m.
High-resolution particle tracking was provided by two
scintillating-fibre trackers (section~\ref{SubSect:Tracker}).
The precise relative alignment of the time-of-flight hodoscopes and
the trackers was obtained by combining the measurements of both
detector systems (section~\ref{SubSect:DA}). 

\graphicspath{{06-Tracking/Figures/}}

\subsection{Trackers}
\label{SubSect:Tracker}

\begin{figure}
  \begin{center}
    \includegraphics[width=0.6\textwidth,keepaspectratio=true,]{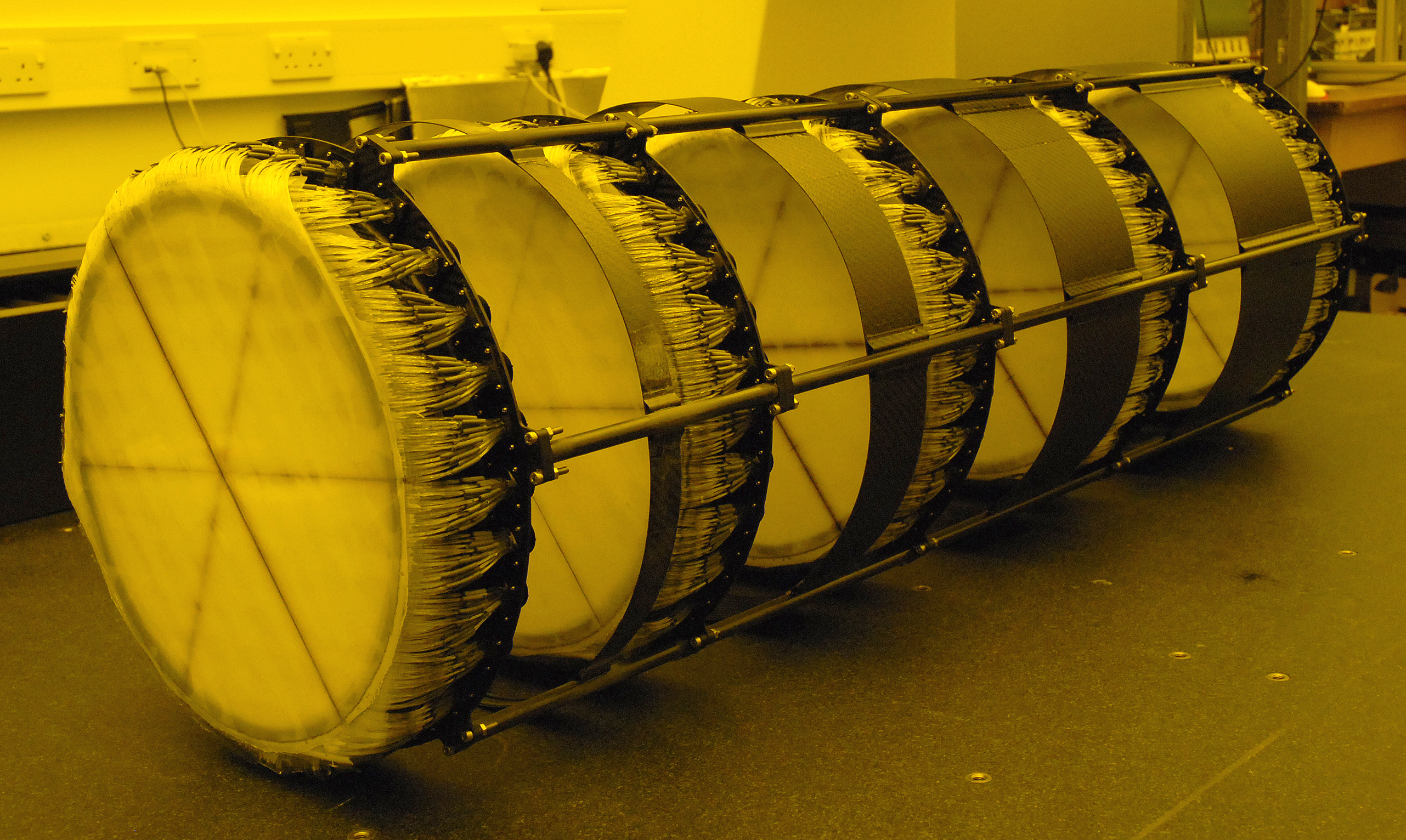}
  \end{center}
  \caption{
    Photograph, with UV-filtered light, of one of the MICE trackers, showing the five stations.
    Each station has three doublet planes of scintillating fibres, each plane
    at 120$^\circ$ to the next (the central fibres of each plane can
    be seen as darker lines traversing the station).
    %Bundles of seven 350\,$\mu$m fibres were grouped together, to be read out by 1\,mm light guides.
  }
  \label{Figure:FullTracker}
\end{figure}

The two high-precision scintillating-fibre trackers each had a
sensitive volume that was 110\,cm in length and 30\,cm in
diameter~\cite{Ellis:2010bb}.
Each tracker was composed of five stations (labelled 1 to 5, with
station 1 being closest to the cooling cell) held in position using a
carbon-fibre space-frame.  
Adjacent stations were separated by different distances ranging from
20\,cm to 35\,cm.
The separations were chosen to ensure that the azimuthal rotation of
track position did not repeat from one station to the next.
This property was exploited in the ambiguity-resolution phase of the
pattern recognition.
Each tracker was instrumented with an internal LED calibration system
and four 3-axis Hall probes to monitor the field.
A photograph of one of the trackers on the bed of the coordinate
measuring machine used to verify the mechanical alignment of the
stations is shown in figure~\ref{Figure:FullTracker}.

Each tracker station consisted of three doublet layers of 350\,$\mu$m
scintillating fibres; these layers were arranged such that each was
set at an angle of 120$^\circ$ with respect to the next.
This arrangement ensured that there were no inactive regions between
adjacent fibres.
Fibres were grouped into one bundle of seven for each readout
channel, to match the resolution to that imposed by multiple
scattering and reduce the overall number of readout channels.
This resulted in a spatial resolution per doublet layer of 470\,$\mu$m
and a measured light yield of approximately 10
photo-electrons~\cite{Ellis:2010bb}.
The light from the seven scintillating fibres was coupled into a
single clear fibre which took it to a visible light photon counter
(VLPC)~\cite{VLPC}.
The signals from the VLPCs were digitised using electronics developed
by the D0 collaboration~\cite{Abazov:2005pn}. \\

\noindent\textbf{Reconstruction} \\
\noindent
The reconstruction software for the trackers is described
in~\cite{Dobbs:2016ejn}.
Each of the 15 doublet layers provided 214 readout channels.
Calibration data taken without beam was used to determine the pedestal and the gain of each channel.
These calibrations were used to correct the number of photoelectrons
(NPE) corresponding to the signal recorded by the tracker
electronics.
The first step in the reconstruction was to record the unique channel
number associated with each NPE value in a ``digit''.
Digit profiles were used to identify hot or dead channels which
were masked from the reconstruction to reduce the rate of ambiguities
that had to be resolved in the pattern recognition and to ensure the
accuracy of the calibration.
The reconstruction proceeded to create ``spacepoints'' from the
intersection of digits in adjacent doublet layers.
Spacepoints were constructed from clusters from all three planes (a
triplet spacepoint) or from any two out of the three planes (a doublet
spacepoint).
The pattern-recognition algorithm searched for spacepoints from
neighbouring stations that were consistent with the helical trajectory
of a charged particle in the solenoidal field.
In the final stage of the tracker reconstruction the track parameters
were estimated using a Kalman filter. \\

\newpage

\noindent\textbf{Noise}\\
\noindent
Digits above a certain NPE threshold were admitted to the
spacepoint-finding algorithm.
Noise in the electronics arising from, for example, the thermal
emission of electrons, could give rise to digits passing the threshold.
Any digit not caused by the passage of a charged particle was
classified as noise.
To isolate noise from signal during beam-on data collection, events
containing a track which included a spacepoint in each of the five
tracker stations were selected.
All digits corresponding to the track were removed from the total set
of digits and the remainder were considered to be noise.
The average noise rate per channel per event was then calculated as
the total number of digits above the NPE threshold divided by the
number of active channels and the  number of events in the sample.
The result of this calculation was that, for an NPE threshold of 2,
the fraction of digits arising from noise was 0.18\% in the upstream
tracker and 0.06\% in the downstream tracker. \\

\noindent\textbf{Track-finding efficiency} \\
\label{trackers:performance:efficiency}
\noindent
The track-finding efficiency was determined using a sample of events
for which the time-of-flight determined from hits in TOF1 and
TOF2 was consistent with passage of a muon.
This requirement ensured that the particle had been transmitted
successfully through the magnetic channel, crossing both trackers.
The track-finding efficiency was defined to be the number of
events in which a track was successfully reconstructed divided by the
total number of events in the sample.
The results of the efficiency analysis are tabulated in
table~\ref{Table:tracker_efficiency_results} for a range of nominal
beam momentum and emittance settings.
The track-finding efficiency obtained in this way averaged over
beam conditions was 98.70\% for the upstream tracker and 98.93\%
for the downstream tracker.
The spacepoint-finding efficiency, defined as the number of
spacepoints found divided by the number of space points expected, was
also determined.
The spacepoint-finding efficiency is summarised for a range of beam
conditions in
table~\ref{Table:tracker_spacepoint_efficiency_results}.
%Overall the high track-finding efficiency was found to be consistent
%with that introduced by the number of dead channels and did not
%introduce significant systematic uncertainties in analysis of the
%MICE data.
\begin{table}
  \caption{
    The track finding efficiency for the upstream and downstream
    trackers for 140\,MeV/$c$ and 200\,MeV/$c$ beams, and for 3, 6 and
    10\,mm nominal emittances.}
  \begin{center}
    \begin{tabular}{| c | c | c | c |}
      \hline 
      \textbf{Momentum} & \textbf{Emittance} & \textbf{Upstream tracks found} & \textbf{Downstream tracks found} \\ \hline
        200 MeV/$c$ & 3\,mm  & 98.38\% & 99.19\% \\ %H36aa
        200 MeV/$c$ & 6\,mm  & 99.42\% & 96.07\% \\ %H25c 
        140 MeV/$c$ & 6\,mm  & 98.37\% & 99.16\% \\ %H36c
        140 MeV/$c$ & 10\,mm & 98.47\% & 98.93\% \\ \hline %H36d
        \multicolumn{2}{| c |}{Average} & 98.70\% & 98.21\% \\
        \hline
    \end{tabular}
  \end{center}
  \label{Table:tracker_efficiency_results}
\end{table}
\begin{table}
  \caption{
    The spacepoint-finding efficiency, in the presence of a
    track, for the upstream and downstream trackers for 140\,MeV/$c$
    and 200\,MeV/$c$ beams, and for 3, 6 and 10\,mm nominal
    emittances.  }
  \begin{center}
    \begin{tabular}{| c | c | c | c |}
      \hline
      \textbf{Momentum} & \textbf{Emittance} & \textbf{Upstream spacepoints found} & \textbf{Downstream spacepoints found} \\ \hline
        200 MeV/$c$ & 3\,mm  & 98.04\% & 97.41\% \\ %H36aa
        200 MeV/$c$ & 6\,mm  & 99.41\% & 94.63\% \\ %H25c 
        140 MeV/$c$ & 6\,mm  & 97.99\% & 99.16\% \\ %H36c
        140 MeV/$c$ & 10\,mm & 98.07\% & 97.44\% \\ \hline %H36d
        \multicolumn{2}{| c |}{Average} & 98.44\% & 97.01\% \\
        \hline
    \end{tabular}
  \end{center}
  \label{Table:tracker_spacepoint_efficiency_results}
\end{table}

\begin{figure}[htb]
  \begin{center}
    \includegraphics[width=0.90\textwidth]{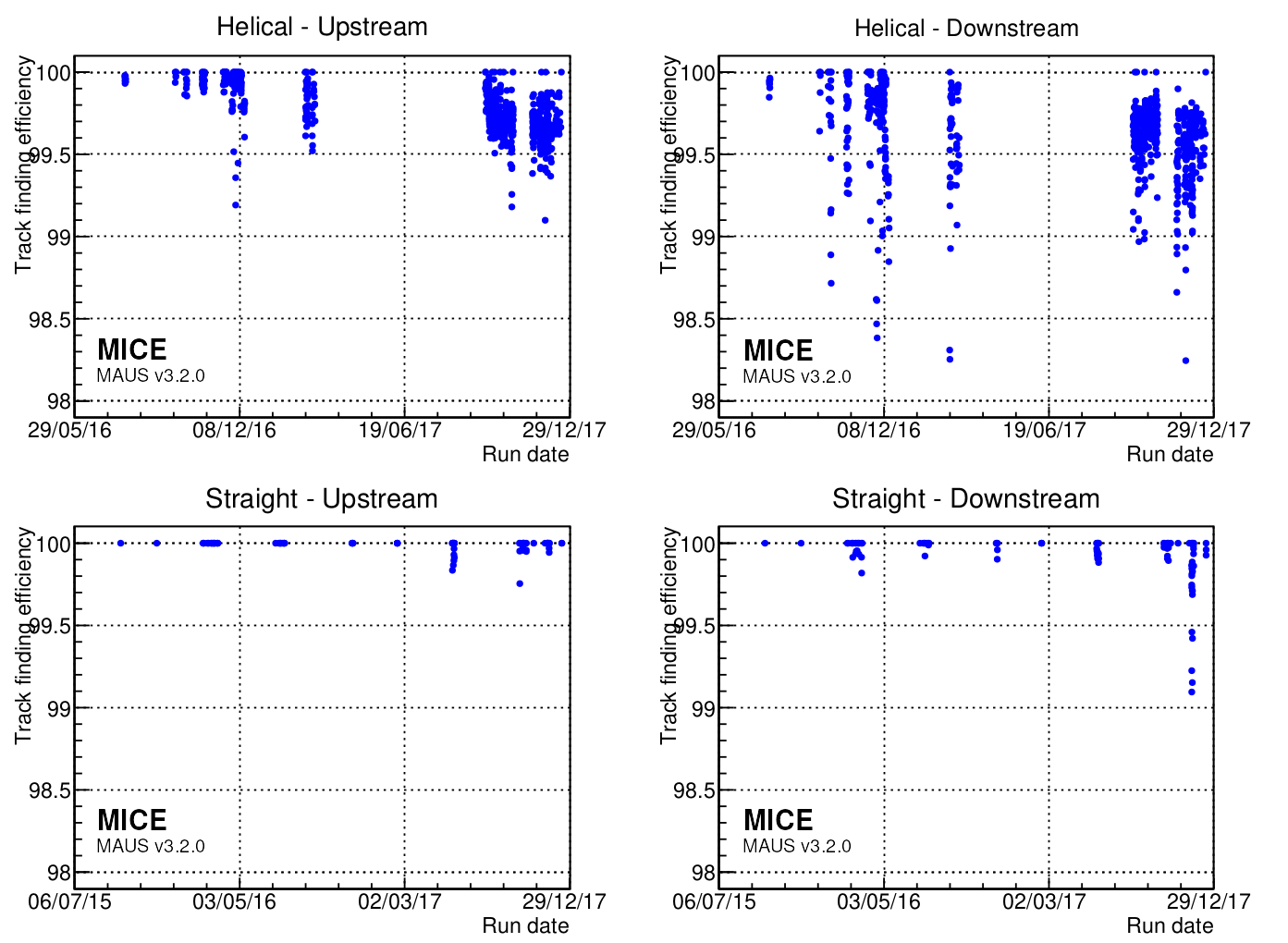}
  \end{center}
  \caption{
    Evolution of the straight and helical track finding efficiencies
    over time for: the upstream (left); and downstream (right) trackers
    during the key periods of data taking since 2015.
    Each dot represents a single data taking run between 10 minutes and 3 hours long.
  }
  \label{fig:trackers:performance:historical} 
\end{figure}

The efficiency of the trackers over the data taking period was
evaluated by selecting events with a measured time-of-flight between
TOF1 and TOF2 consistent with the passage of a muon.
Events were required to contain at least one hit within the fiducial
volume of the tracker.
An event was added to the numerator of the efficiency calculation if
it contained a single space point in each of the five tracker
stations.
The evolution of the tracking efficiency in the upstream and
downstream trackers is shown in
figure~\ref{fig:trackers:performance:historical}.
The efficiency is shown separately for data taken in the presence of a
magnetic field (``helical'') and with the solenoids turned off
(``straight'').
The data shows that the efficiency was generally greater than 99.0\%.
Water vapour ingress to the cold end of the VLPC cassettes caused the loss of
channels and contributed to a reduction in the tracking efficiency.
This was recovered by warming and drying the VLPCs. \\

\begin{figure}[htb]
  \begin{center}
    \begin{tabular}{cc}
      \includegraphics[width=0.49\textwidth]{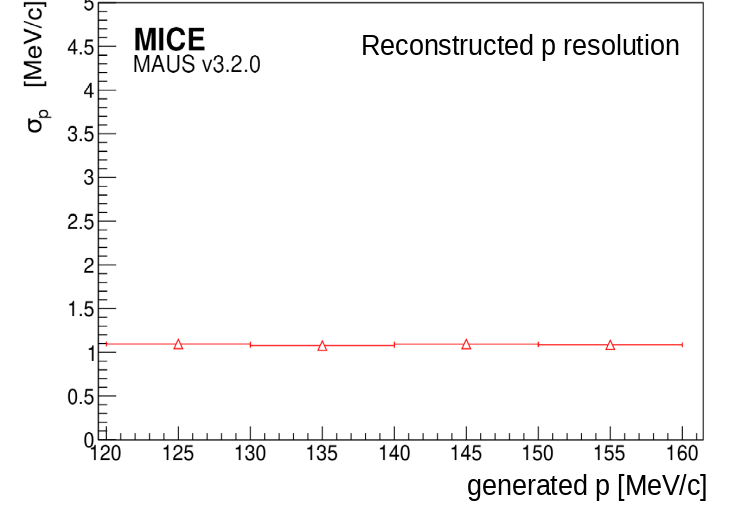} &	
      \includegraphics[width=0.49\textwidth]{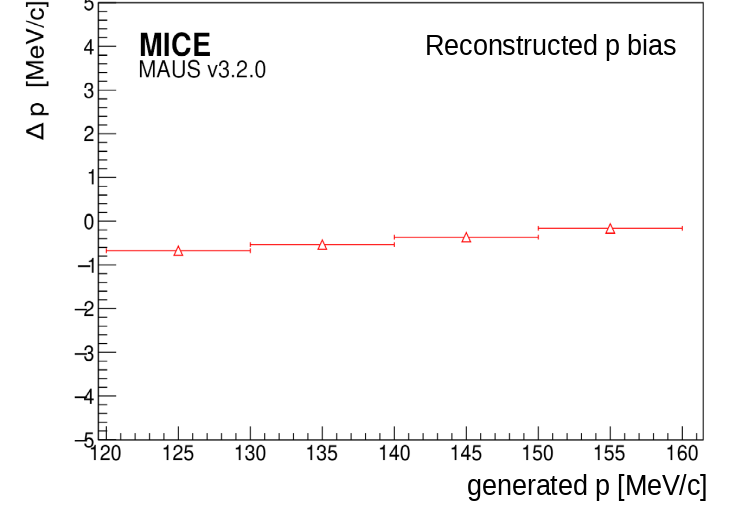} \\
      \includegraphics[width=0.49\textwidth]{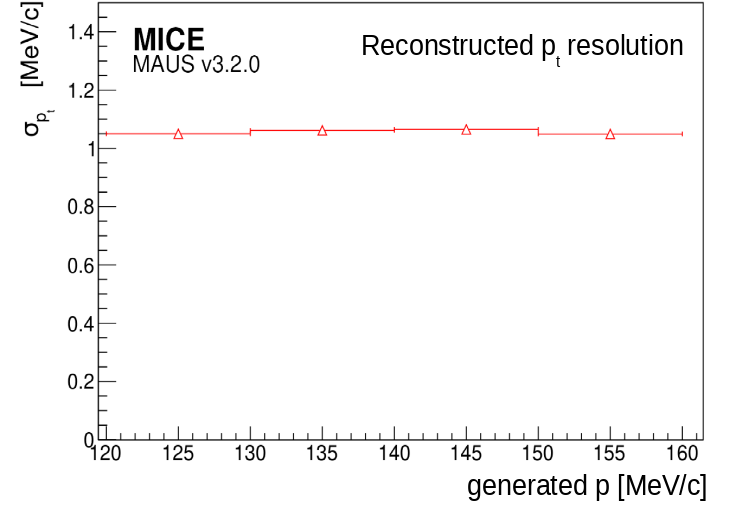} &
      \includegraphics[width=0.49\textwidth]{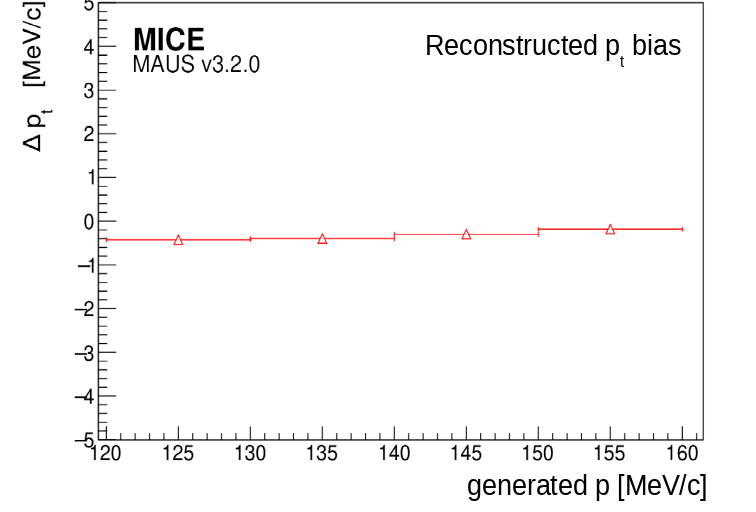}
    \end{tabular}
  \end{center}
  \caption{
    %Upstream tracker: from simulated data.
    Momentum reconstruction resolution (left) and bias
    (right) for the total momentum (top) and transverse momentum
    component (bottom) in the upstream tracker.
  }
  \label{trackers:performance:resolutions:up}
\end{figure}

\begin{figure}[htb]
  \begin{center}
    \begin{tabular}{cc}
      \includegraphics[width=0.49\textwidth]{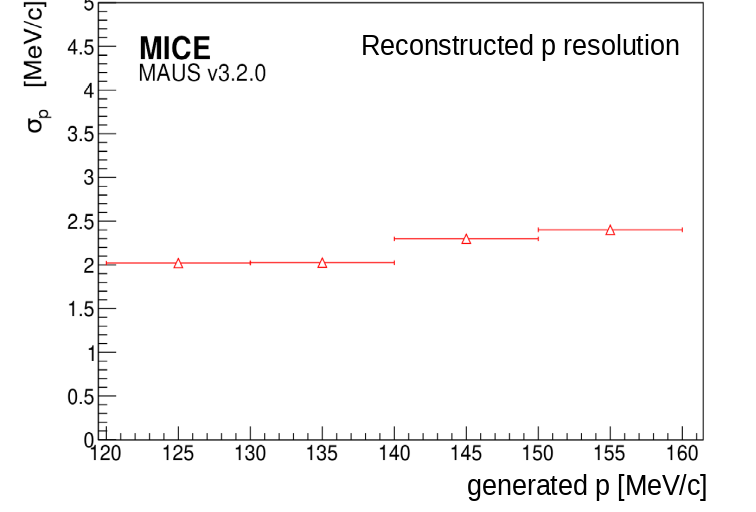} &	
      \includegraphics[width=0.49\textwidth]{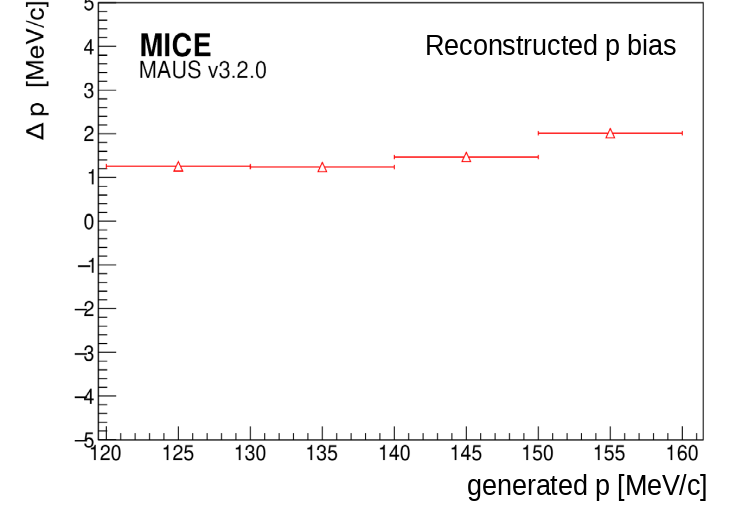} \\
      \includegraphics[width=0.49\textwidth]{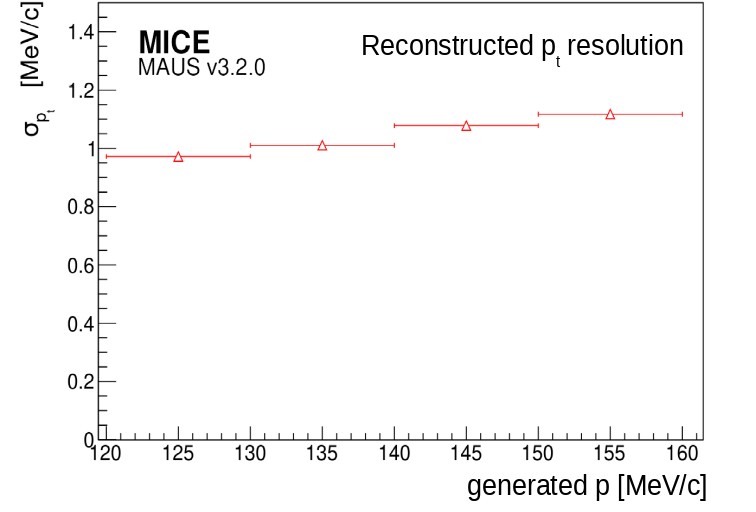} &
      \includegraphics[width=0.49\textwidth]{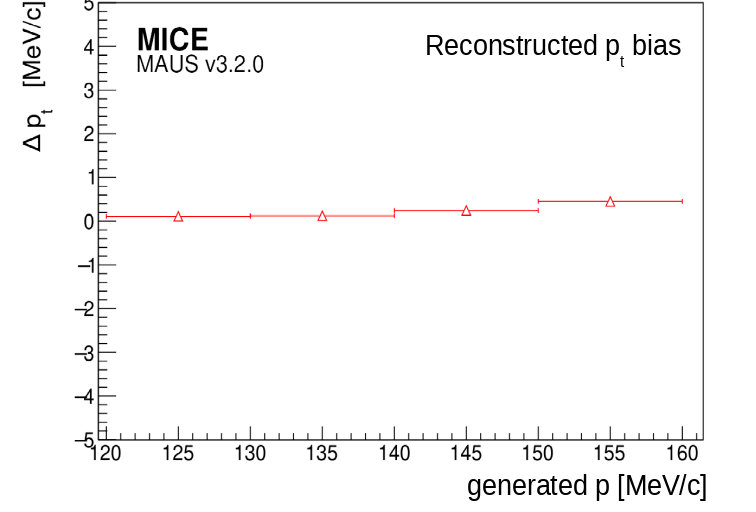}
    \end{tabular}
  \end{center}
  \caption{
    %Upstream tracker: from simulated data.
    Momentum reconstruction resolution (left) and
    bias (right) for the total momentum (top) and transverse
    momentum component (bottom) in the downstream tracker.
  }
  \label{trackers:performance:resolutions:down}
\end{figure}

\noindent\textbf{Track-fit performance} \\
\noindent
Monte Carlo simulation with realistic field, beam conditions and detector geometry was used to estimate the performance of the track fit.
A beam centred at 140\,MeV/$c$ with 10\,mm nominal emittance,
representing a typical data set, was used for the study.
Results are presented in
figure~\ref{trackers:performance:resolutions:up} for the upstream
tracker and figure~\ref{trackers:performance:resolutions:down} for the
downstream tracker.
The resolution in the total momentum and transverse momentum is
observed to be $\sim1.1$\,MeV/$c$ independent of momentum in the range
120\,MeV/$c$ to 160\,MeV/$c$.
%The small bias in reconstructed momentum and transverse momentum is {\color{red} understood to be due to ...?}
The small bias in the transverse and the total momentum did not give rise to significant effects in the analysis and was considered in systematic error studies.

%\graphicspath{{09-Detector-alignment/Figures/}}

\subsection{Beam-based detector alignment}
\label{SubSect:DA}

A beam-based alignment algorithm was developed to improve the
resolution on the position of the scintillating-fibre trackers
relative to the time-of-flight hodoscopes.
The starting point for the beam-based alignment was the geometrical
survey of the detectors in the MICE Hall which was performed using
laser geodesy. 
Survey monuments on the TOF frames were surveyed with respect to the
MICE Hall survey network.
The trackers had been dowelled in position in the bores of the
spectrometer solenoids.
The dowels were used to locate each tracker precisely with respect to the axis of the warm bore of its
solenoid.
The position of the trackers along the beam line was inferred from the
measurement of survey monuments mounted on the spectrometer-solenoid
cryostats outer jackets.
The beam-based alignment was used to determine the azimuthal
orientation of the trackers with a resolution of 6\,mrad/$\sqrt{N}$
and their position transverse to the beamline with a resolution of
20\,mm/$\sqrt{N}$, where $N$ is the number of tracks used in the
analysis~\cite{2018arXiv1805.06623T}. \\

\noindent\textbf{Analysis method} \\
\label{SubSect:DA_Analysis}
\noindent The position of each tracker in the MICE Hall coordinate
system was described using the location of its centre and a set of
three angles corresponding to rotation about the $x$ axis ($\alpha$),
the $y$ axis ($\beta$) and the $z$ axis ($\phi$).
The rotation of the tracker about the $z$ axis has a negligible effect
on the alignment since $\phi$ was determined precisely
at installation.
An initial estimate for the position of each tracker along the beamline
had been inferred from the survey.
The surveyed location of the TOFs was used as the reference for the
tracker alignment.
The line that joins the centre of TOF1 with the centre of TOF2 was
chosen as the reference axis.
A deviation from this axis was considered to be due to misalignment
of the trackers.
The alignment could not be determined on a single-particle basis due
to multiple Coulomb scattering in the absorber and other material
present on the beamline.
Therefore, the mean residuals in position ($x$ and $y$) and angle
($\alpha$ and $\beta$) of the trackers with respect to the TOF1--TOF2
axis were evaluated to determine the alignment constants.

Each TOF provided a single spacepoint in the Hall coordinate system.
In Hall coordinates, on average, the track reconstructed between
TOF1 and TOF2 should agree with the track reconstructed in each 
tracker, i.e. the mean residuals in $x, y, \alpha$, and $\beta$
should be zero. 
Applying this reasoning to the unknown offset and angles leads to a
system of equations for the four unknown
constants~\cite{2018arXiv1805.06623T}.
The measurement of four residual distributions per tracker yields the
alignment constants. 
The main source of bias was the scattering in the material between
TOF1 and TOF2.
If the beam was not perfectly centred, particles preferentially
scraped out on one side of the magnet bore, anisotropically truncating
the tail of the residual distribution. 
A fiducial cut was applied to the upstream sample in order to remove
this effect.

\begin{figure}[htb]
  \begin{center}
    \begin{minipage}[b]{.45\textwidth}
      \begin{center}
        \includegraphics[width=\textwidth]{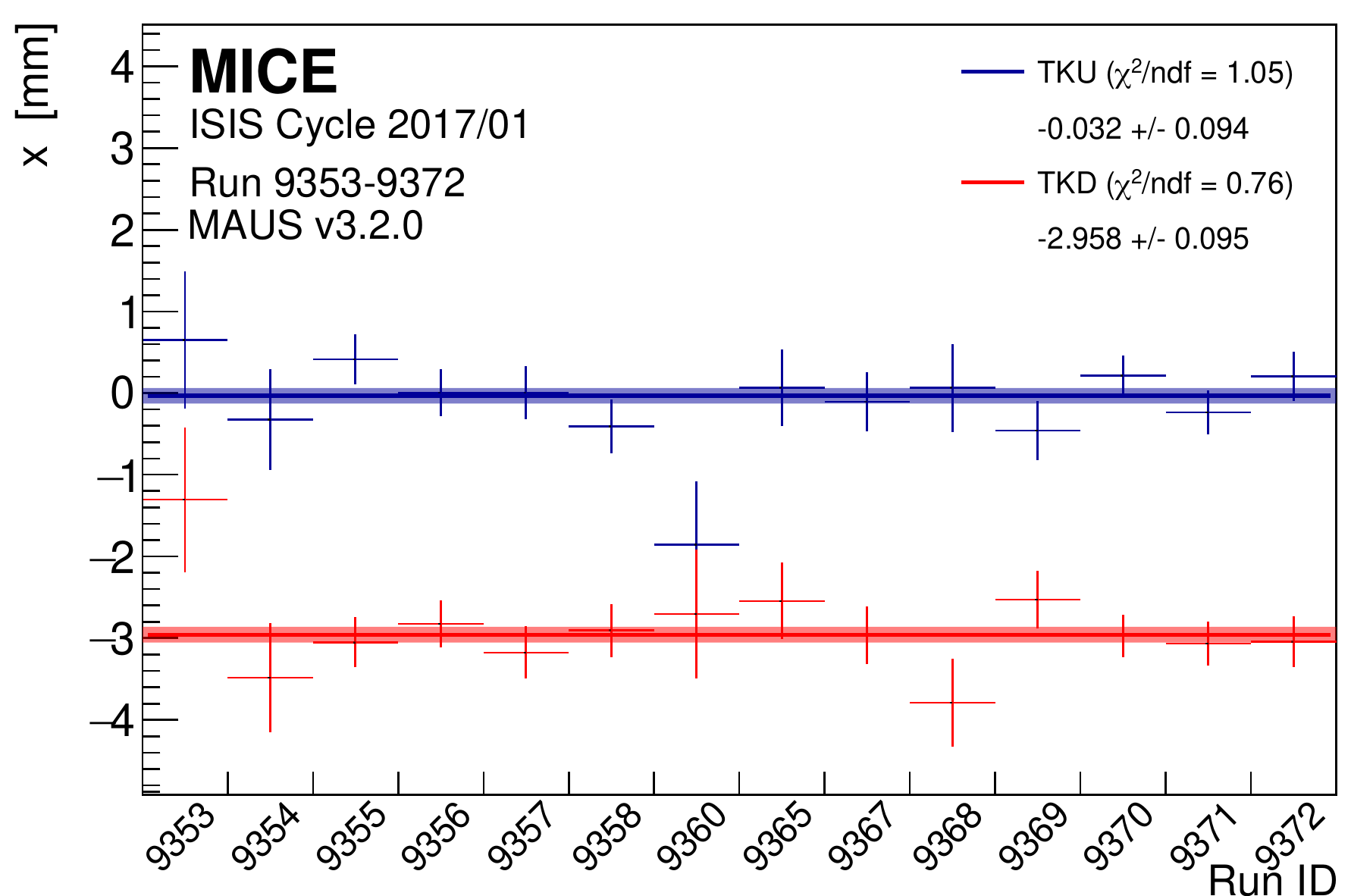}
      \end{center}
    \end{minipage}
    \hfill
    \begin{minipage}[b]{.45\textwidth}
      \begin{center}
        \includegraphics[width=\textwidth]{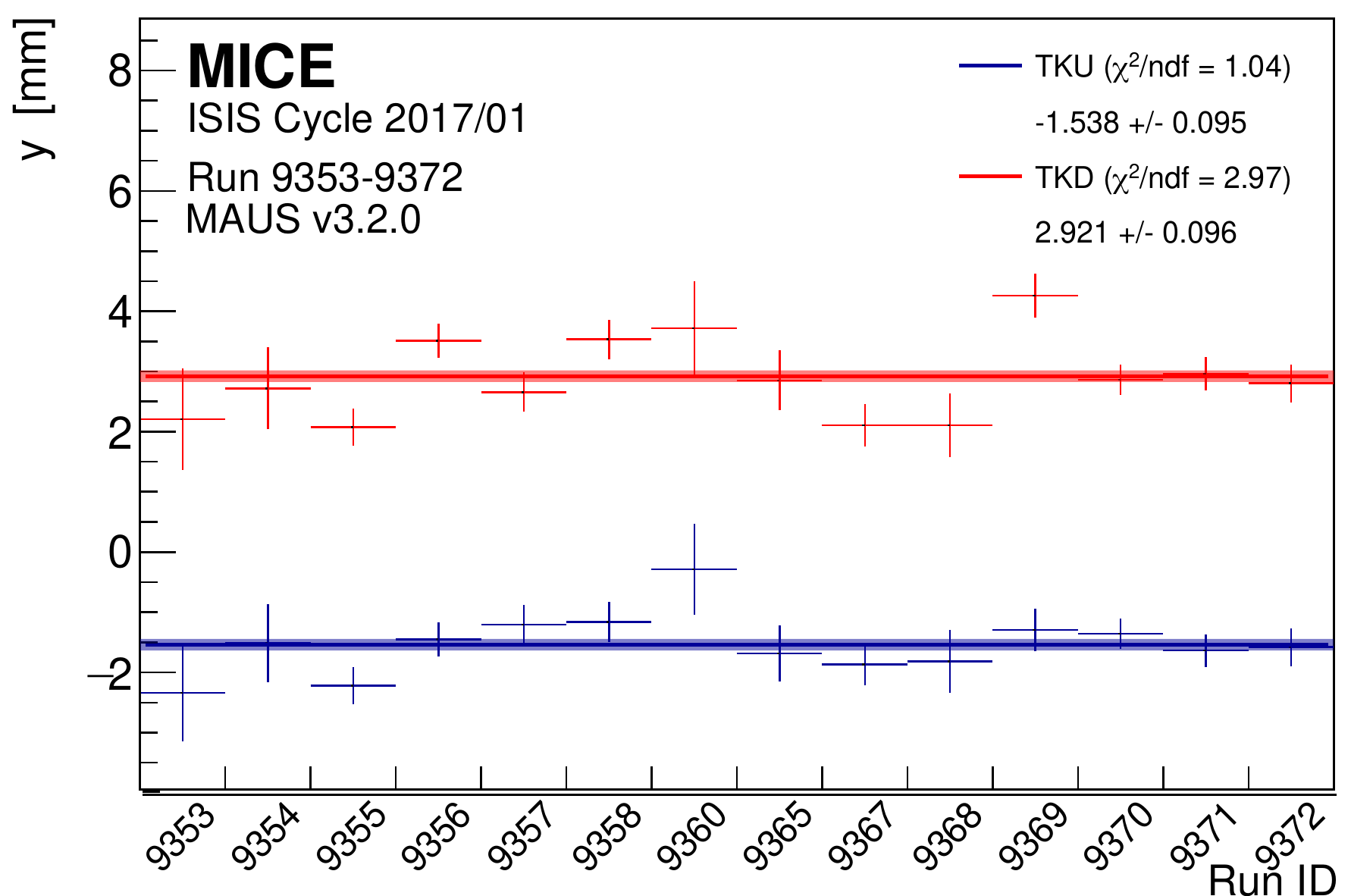}
      \end{center}
    \end{minipage}
    \begin{minipage}[b]{.45\textwidth}
      \begin{center}
        \includegraphics[width=\textwidth]{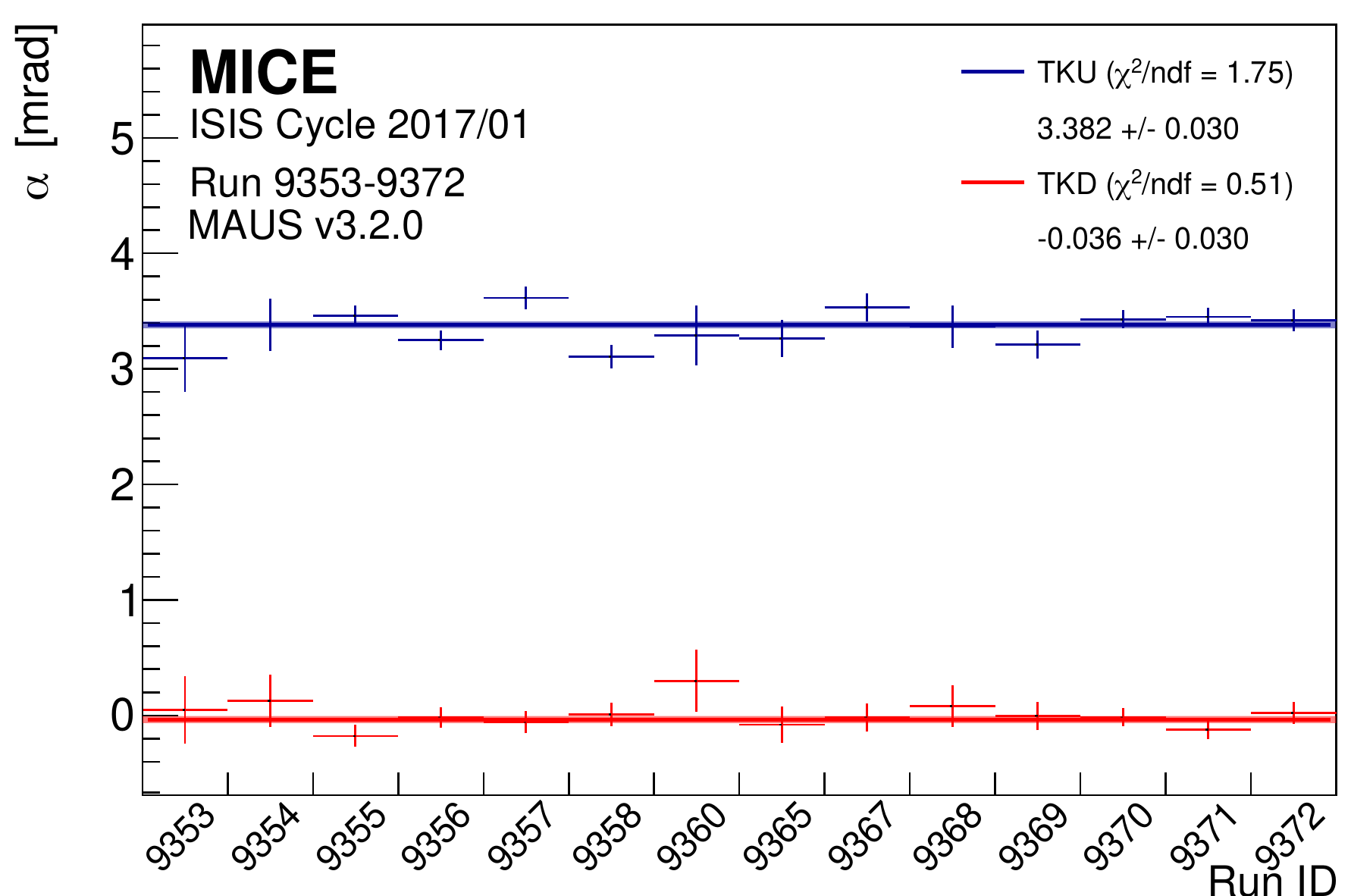}
      \end{center}
    \end{minipage}
    \hfill
    \begin{minipage}[b]{.45\textwidth}
      \begin{center}
        \includegraphics[width=\textwidth]{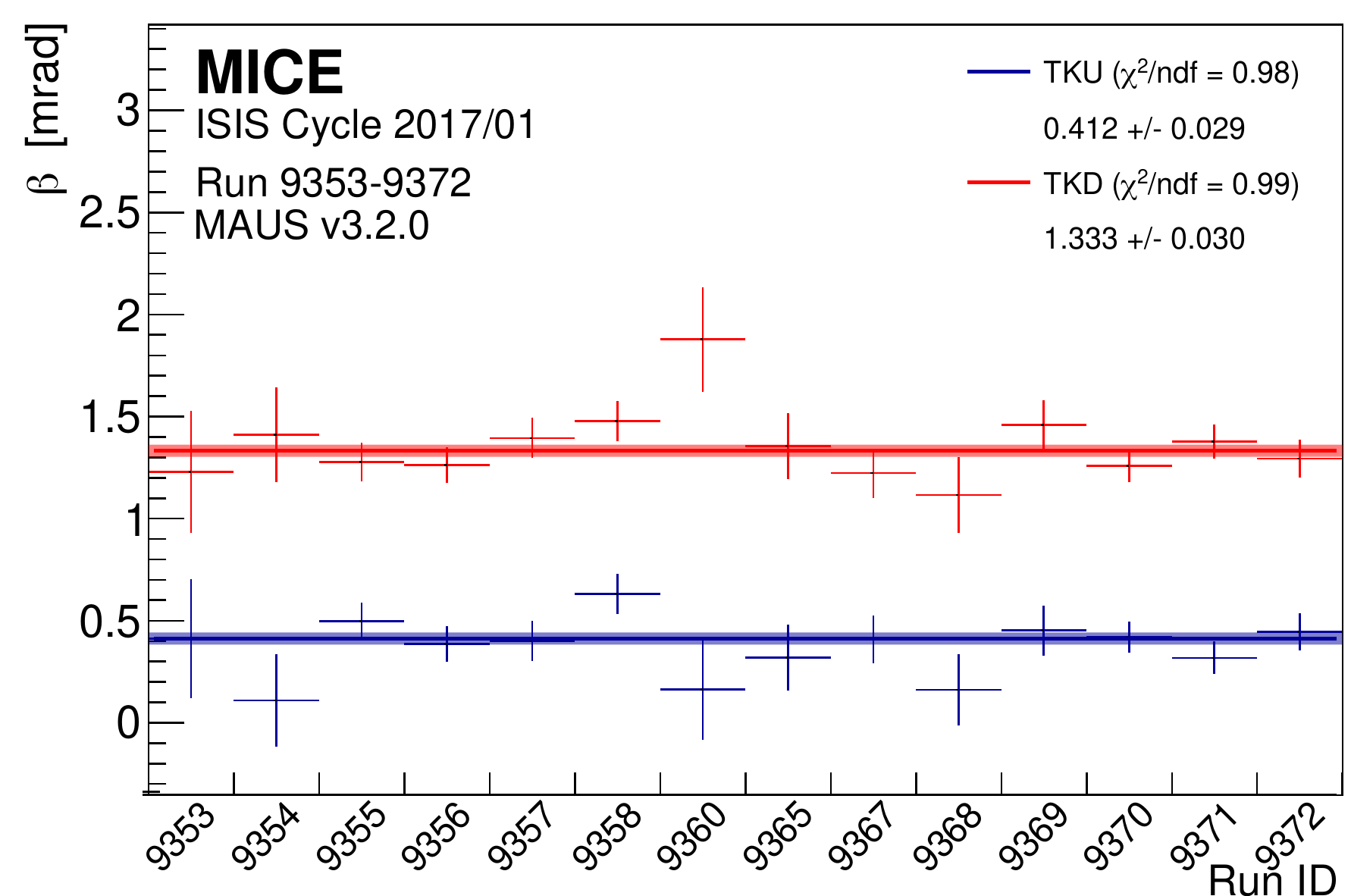}
      \end{center}
    \end{minipage}
  \end{center}
  \caption{
    Consistency of the alignment algorithm results for upstream (blue) and downstream (red) trackers  across runs acquired during
    the 2017/01 ISIS user cycle. The quantities $x$, $y$, $\alpha$, and $\beta$ are defined in the text.
  }
  \label{fig:runtorun}
\end{figure}

Data were recorded with the superconducting magnets turned off.
High momentum beams were used to reduce the RMS scattering angle and
to maximise transmission.   
Each data set was processed independently.
Figure~\ref{fig:runtorun} shows the alignment parameters determined for
each run during a specific data taking period.
The measurements are in good agreement with one another and show no
significant discrepancy: an agreement between the independent fits
guaranteed an unbiased measurement of the alignment constants.
The constant-fit $\chi^2/\text{ndf}$ was close to unity for each fit,
indicating that there were no additional sources of significant
uncertainty.
The optimal parameters are summarised in
table~\ref{tab:201701_constants}. 
\begin{table}
  \caption{
    Optimal alignment constants measured in the high-momentum straight-track data
    acquired during May 2017 (summarised from figure~\ref{fig:runtorun}).
    }
    \begin{center}
    \begin{tabular}{|l|c|c|c|c|}
  \hline
	& \textbf{x [mm]} & \textbf{y [mm]} & \textbf{$\alpha$ [mrad]} & \textbf{$\beta$ [mrad]} \\
	\hline
	\textbf{TKU} & $-0.032\pm0.094$ & $-1.538\pm0.095$ & $ 3.382\pm0.030$ & $0.412\pm0.029$ \\
	\textbf{TKD} & $-2.958\pm0.095$ & $ 2.921\pm0.096$ & $-0.036\pm0.030$ & $1.333\pm0.030$ \\
	\hline
    \end{tabular}
  \end{center}
  \label{tab:201701_constants}
\end{table}

\graphicspath{{07-Absorber/Figures/}}

\section{Liquid Hydrogen Absorber}
\label{Sect:Absorber}

The accurate characterisation of the properties of the liquid hydrogen
absorber was a critically-important contribution to the study of
ionisation cooling.
The instrumentation used for this purpose and its performance are
presented in this section.

\begin{figure}
  \begin{center}
    \includegraphics[width=0.80\textwidth]{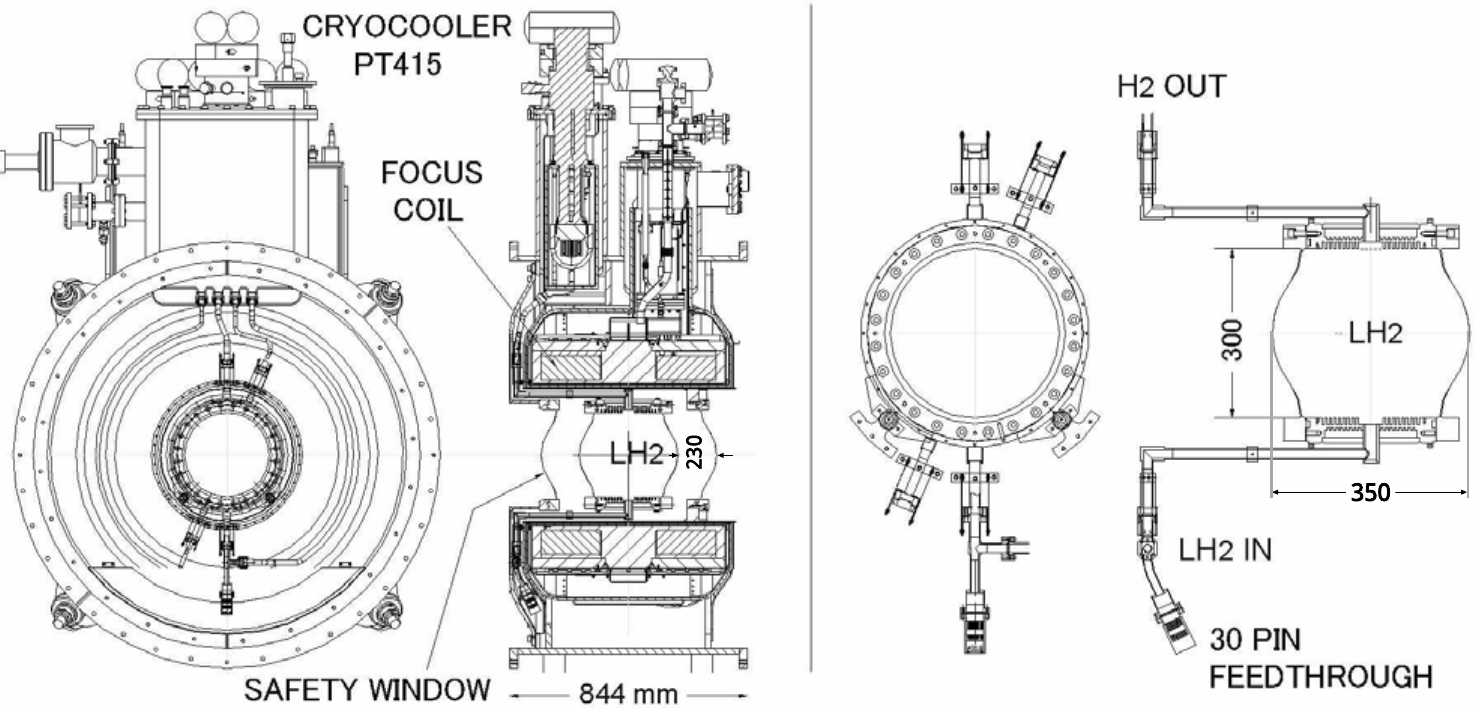}
  \end{center}
  \caption{
    Left panel: Drawing of the focus coil (FC) module
    showing the principal components.
    Right panel: detail of the liquid hydrogen absorber vessel~\cite{1748-0221-13-09-T09008}.
  }
  \label{Fig:AbsorberVessel:Diag}
\end{figure}

The absorber vessel consisted of a cylindrical aluminium body sealed
with two thin aluminium end windows, as shown in
figure~\ref{Fig:AbsorberVessel:Diag}.
The absorber vessel contained 22\,l of liquid.
The body of the absorber had an inner diameter of 300\,mm and the end
flanges were separated by a distance of 230\,mm.
The vessel was surrounded by a second pair of safety
windows.
The length along the central axis, between the
two domes of the end windows, was 350\,mm~\cite{1748-0221-13-09-T09008}.
 \\

\noindent\textbf{Variation of the density of liquid hydrogen due to
    varying temperature and pressure} \\
\noindent
The energy lost by a muon travelling through the liquid hydrogen
absorber depends on the path length and on
the density of the liquid hydrogen. The density of liquid hydrogen is
a function of temperature and pressure.  
The temperature of the vessel was measured by eight LakeShore Cernox
1050 SD sensors, but with the values truncated for storage
at a granularity of 0.1\,K.
Four of the sensors were used solely as temperature sensors, while the
other four were also used as level sensors to ensure the
liquid hydrogen reached the top of the vessel. 
The sensors were arranged in pairs, with two mechanically clamped at
the top of the vessel, two at a polar angle of ${45}^{\circ}$ to
vertical from the top of the vessel, two at a polar angle of
${45}^{\circ}$ to the bottom of the vessel, and a
final two at the bottom of the vessel. 

Cooldown and liquefaction were completed slowly over eight days at a
pressure of 1105\,mbar after which the vessel's pressure was lowered to
1085\,mbar~\cite{1748-0221-13-09-T09008}.
The vessel then remained in this steady state during the 21 day period of data taking, after which the vessel was vented. For the venting process,
the cryocooler used to liquefy hydrogen was
switched off and heaters were switched on to deliver a nominal power
of 50\,W to the absorber vessel.
This resulted in an increase in pressure to 1505\,mbar until the
temperature stabilised at the boiling point.
A rapid increase in temperature was observed once all the
liquid hydrogen had boiled off. 

The temperature sensors had a typical accuracy of
$\mathrm{\pm}$\,9\,mK and a long-term stability of
$\mathrm{\pm}$\,12\,mK at 20\,K.
The magnetic-field dependent temperature error, $\Delta$\textit{T}/\textit{T}, at 2.5\,T is 0.04\%,
equivalent to $\mathrm{\pm}$\,8\,mK at
20\,K~\cite{TemperatureMeasurement}.
These uncertainties were quoted by the manufacturer of the sensors.
Magnetic fields caused reversible calibration shifts on the temperature
measurements.
To reduce the uncertainty in the liquid hydrogen density a calibration
procedure was devised that used the boiling point, as observed
during the venting process.
A correction to the observed temperature reading was obtained by
applying a cut-off correction, a correction for the effect of the
magnetic field based on the current in the focus coil and its
polarity, a correction for the non-linearity of the sensors, and a 
boiling point scaling factor~\cite{NOTE524}.  
 
The boiling point of hydrogen at 1085\,mbar is 20.511\,K.
The sensors had a total uncertainty of 17\,mK (9\,mK accuracy, 12\,mK
stability, 8\,mK magnetic).
The deviation from the non-linearity of the sensors~\cite{TemperatureMeasurement} added, on average,
0.03\,K to the uncertainty.
The temperature scaling and magnet-current correction factors also
had an associated uncertainty as
they were derived based on the 0.1\,K resolution of the retrieved, truncated, values.
For example, a calibrated sensor at boiling temperature and 1505\,mbar
should read 21.692\,K, but we can only retrieve a value of
21.65\,K (21.6\,K truncated plus 0.05 K cut-off correction~\cite{NOTE524}) i.e.
off by 0.042\,K.
The pressure sensors had an uncertainty of $\mathrm{\pm}$\,5\,mbar
which equated to $\mathrm{\pm}$\,0.016\,K during steady state.
The pressure uncertainty ($\mathrm{\pm}$\,5\,mbar) added another
uncertainty to the temperature calibration constants of
$\mathrm{\pm}$\,0.014\,K.
Collectively, all these uncertainties summed in quadrature to 0.2\,K for
each sensor.
 
\begin{figure}
  \begin{center}
    \includegraphics[width=0.80\textwidth]{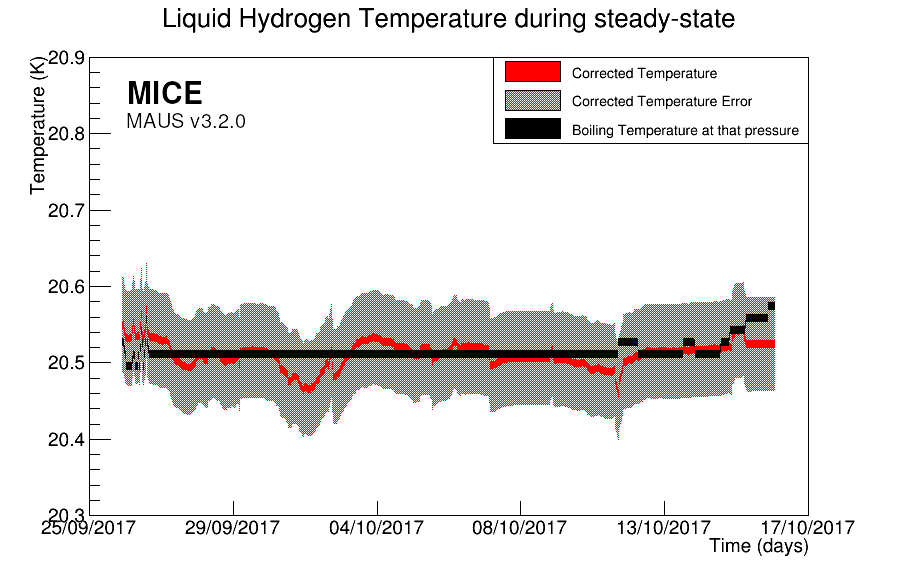}
  \end{center}
  \caption{
    Average liquid hydrogen temperature recorded by the sensors during the
    steady state period.
    After applying all the correction factors the temperature remains
    at or close to the boiling point temperature.
  }
  \label{Fig:TempCalibrated}
\end{figure}

While in the steady state condition the liquid hydrogen was close to the
boiling temperature of liquid parahydrogen~\cite{NOTE524} (density of 70.53\,kg/m$^{3}$):
the average temperature of the eight sensors was (20.51\,$\mathrm{\pm}$\,0.07)\,K at 1085\,mbar
(figure~\ref{Fig:TempCalibrated}) allowing us to determine the
uncertainty in the density over this period as 0.08\,kg/m$^{3}$. \\

\noindent\textbf{Contraction of the absorber vessel due to cooling} \\
\noindent
The absorber was cooled from room temperature to the operating
temperature of the experiment (20.51\,K), contracting the vessel.
The linear contraction of Al-6061 as it is cooled from 293\,K is given
by: 
\begin{equation}
  \alpha =-4.1277\times {10}^{-3}T-3.0389\times {10}^{-6}T^2+8.7696\times {10}^{-8}T^3-9.9821\times {10}^{-11}T^4
\end{equation}
where $T$ is the operating temperature~\cite{Hardin}.
The equation is the result of a fit to data collated by the National
Institute of Standards and Technology (NIST) and has an associated
curve fit error of 4\%. 
At the MICE operating temperature, this corresponds to a linear
contraction of the vessel along each plane of 0.415\%.
As a result the length of the bore contracted by
$(1.45 \pm 0.05)$\,mm.
The vessel was suspended within the warm bore of the focus coil and
was therefore free to contract in each plane without restriction.  \\

\noindent\textbf{Deflection of absorber vessel windows due to internal
  pressure} \\
\noindent
To minimise energy loss and Coulomb scattering by the absorber vessel,
the window thickness was minimised.
The liquid hydrogen circuit was pressurised
above atmospheric pressure to prevent air ingress~\cite{1748-0221-13-09-T09008,Ishimoto}. 
The vessel was designed to withstand at least 2500\,mbar internally.
The internal pressure was limited by the 1.5\,bar relief valve to atmosphere, whilst the vessel was surrounded by vacuum.

The pressure at which the absorber operated resulted in deflection of the absorber windows. These
deflections were modelled using ANSYS~\cite{NOTE155}, and the uncertainty in the
window deflection derived from this model was 20\%.
The model showed a linear dependence of the window deflection on
pressure up to 2\,Bar when the windows begin to yield.
The pressure sensors were accurate to $\mathrm{\pm}$\,5\,mbar
(0.25\% of 2\,Bar).
At (1085\,$\mathrm{\pm}$\,5)\,mbar, the typical MICE operating
pressure, this corresponded to a deflection of
(0.5374\,$\mathrm{\pm}$\,0.1076)\,mm (model uncertainty)
$\mathrm{\pm}$\,0.0022\,mm (sensor uncertainty) at the centre of the
absorber window. \\

\noindent\textbf{Variation of the absorber vessel window thicknesses} \\
\noindent
On its passage through the absorber a muon would lose energy in the
aluminium of the pair of hydrogen-containment windows, the two
aluminium safety windows, and the liquid hydrogen itself.
At the centre of the absorber, the total amount of aluminium the muon
beam passed through was (785\,$\mathrm{\pm}$\,13)\,$\mu$m, producing a variance
of 1.68\%.
However, as the windows were thin, the effects on energy loss were
negligible.
A 200\,MeV/$c$ muon passing along the central axis of an empty
absorber lost 0.345\,MeV, introducing a 0.006\,MeV uncertainty
on energy loss.  \\

\noindent\textbf{Total systematic uncertainty on energy loss} \\
\noindent
The principal contributions to the systematic uncertainty on energy
loss in the liquid hydrogen absorber are: the uncertainty in the
contraction of the absorber vessel, the uncertainty in the deflection
of the hydrogen-containment windows due to internal pressure, and the
uncertainty in the variation of the window thickness.
The impact of the contraction of vessel and the deflection of the
windows resulted in a reduction of the length of the vessel on
axis of (0.4\,$\mathrm{\pm}$\,0.2)\,mm.
The change in the combined thicknesses of the absorber
windows on axis is 13\,$\mu$m.
The average temperature during the steady state period of the
experiment when the pressure remained constant at
(1085\,$\mathrm{\pm}$\,5)\,mbar is (20.51\,$\mathrm{\pm}$\,0.07)\,K 
corresponding to a liquid hydrogen density of (70.53\,$\mathrm{\pm}$\,0.08)\,kg/m$^{3}$.

During the MICE data taking, muon beams with nominal momenta of 140,
170, 200 and 240\,MeV/$c$ were used.
The energy loss and its uncertainty were calculated.
The calculation used a central bore length of
(349.6\,$\mathrm{\pm}$\,0.2)\,mm, a total window thickness of
(0.785\,$\mathrm{\pm}$\,0.013)\,mm and a liquid hydrogen density of
\linebreak[4] 
(70.53\,$\mathrm{\pm}$\,0.08)\,kg/m$^{3}$.
For a 140\,MeV/$c$ muon this corresponds to an energy loss of
(10.88\,$\mathrm{\pm}$\,0.02)\,MeV, while for a 200\,MeV/$c$ muon 
particle this corresponds to an energy loss of
(10.44\,$\mathrm{\pm}$\,0.02)\,MeV.
For a muon travelling along the centre axis of the absorber the
systematic uncertainty in the energy loss is 0.2\%.

\graphicspath{{80-Conclusions/Figures/}}

\section{Summary and conclusions}
\label{Sect:Conclusions}

A complete set of particle detectors has permitted the full characterisation and study of the evolution of the phase space of a muon beam through a section of a cooling channel in the presence of liquid hydrogen and lithium hydride absorbers, leading to the first measurement of ionization cooling.
The PID performance of the detectors is summarised in table~\ref{tab:pid1} and table~\ref{tab:pid2} and is fully compatible with the specification of the apparatus~\cite{NOTE21}.

\begin{table}[htb!]
	\caption{Summary of the performance of the MICE PID detectors.}
  \begin{center}
	\begin{tabular}{|c|c|c|}
   	\hline
	  \textbf{Detector}              & \textbf{Characteristic}            & \textbf{Performance} \\
		\hline
    Time-of-Flight        & time resolution           & 0.10\,ns    \\
%   Cherenkov             & ??                        & ??          \\
    KLOE-Light            & muon PID efficiency       & ~99\%       \\
    Electron Muon Ranger  & electron PID efficiency   & 98.6\%      \\
%   Trackers              & track finding efficiency  & $>$98\%     \\
    \hline
  \end{tabular}
	\label{tab:pid1}
  \end{center}
\end{table}

\begin{table}[htb!]
	\caption{Summary of the MICE PID detector performance for different beam settings.}
  \begin{center}
  \begin{tabular}{|c|ccc|cc|cc|cc|cc|}
    \hline
  &

  \multicolumn{3}{c|}{\textbf{KL efficiency}} &
  \multicolumn{2}{c|}{\textbf{EMR efficiency}} &
  \multicolumn{6}{c|}{\textbf{Track finding efficiency}} \\ 
  \hline
  \multirow{2}{*}{Momentum} &
  \multirow{2}{*}{electrons} &
  \multirow{2}{*}{muons} &
  \multirow{2}{*}{pions} &
  \multirow{2}{*}{electrons} &
  \multirow{2}{*}{muons} &
  \multicolumn{2}{c|}{3 mm} &
  \multicolumn{2}{c|}{6 mm} &
  \multicolumn{2}{c|}{10 mm} \\  
             &      &      &      &       &      & US   & DS   & US   & DS   & US   & DS   \\ \hline
\textbf{140\,MeV/$c$} & 95\% & 97\% & n.a.   & 98\%  & 35\% &        &      & 98\%   & 99\% & 98\% & 99\% \\ \hline
\textbf{170\,MeV/$c$} & 95\% & 99\% & 89\% & 99\%  & 99\% &        &      &          &        &        &      \\ \hline
\textbf{200\,MeV/$c$} & 94\% & 99\% & 95\% & 100\% & 99\% & 99\% & 96\% & 99\% & 96\% &        &      \\ \hline
\textbf{240\,MeV/$c$} & 96\% & 99\% & 97\% & 99\%  & 99\% &        &      &          &        &        &      \\ \hline
\textbf{300\,MeV/$c$} & 95\% & 99\% & 98\% & n.a.    & 99\% &        &      &          &        &        &      \\ \hline
  \end{tabular}
	\label{tab:pid2}
  \end{center}
\end{table}
All the different elements of the MICE instrumentation have been used to characterise the beam and the measurement of the cooling performance for a different variety of beam momenta, emittance, and absorbers. The measurement of the physical properties of the liquid hydrogen absorber have been fully described here.
The experiment has thus demonstrated a technique critical for a muon collider and a neutrino factory and brings those facilities one step closer.

\section{Acknowledgements}
\label{Sect:Acknowledgements}
The work described here was made possible by grants from the Department of Energy and National Science Foundation (USA), the Istituto Nazionale di Fisica Nucleare (Italy), 
the Science and Technology Facilities Council (UK), the European Community under the European Commission Framework Programme 7, the Japan Society for the Promotion of 
Science and the Swiss National Science Foundation, in the framework of the SCOPES programme. We gratefully acknowledge all sources of support.
We acknowledge the use of Grid computing resources deployed and operated by GridPP in the UK~\cite{grid_pp_2009}. We are also grateful to the staff of ISIS for the reliable operation of ISIS.

The MAUS software used to reconstruct and analyse the MICE data is available at~\cite{MICE_code}.

\clearpage
\bibliographystyle{99-Styles/utphys}
\bibliography{Concatenated-bibliography}

\newpage
\appendix
%======= Author list: 2021, System performance paper
%

\thispagestyle{plain}
\setlength\parindent{0em}%\noindent

\section*{The MICE collaboration}

M.~Bogomilov,  R.~Tsenov, G.~Vankova-Kirilova
\\{\it
Department of Atomic Physics, St.~Kliment Ohridski University of Sofia, 5 James Bourchier Blvd, Sofia, Bulgaria
}\\

Y.~P.~Song, J.~Y.~Tang
\\{\it
Institute of High Energy Physics, Chinese Academy of Sciences, 19 Yuquan Rd, Shijingshan District, Beijing, China
}\\

Z.~H.~Li
\\{\it
Sichuan University, 252 Shuncheng St, Chengdu, China
}\\

R.~Bertoni, M.~Bonesini, F.~Chignoli, R.~Mazza
\\{\it
Sezione INFN Milano Bicocca and Dipartimento di Fisica G.~Occhialini, Piazza della Scienza 3, Milano, Italy
}\\

V.~Palladino
\\{\it
Sezione INFN Napoli and Dipartimento di Fisica, Universit\`{a} Federico II, Complesso Universitario di Monte S.~Angelo, via Cintia, Napoli, Italy
}\\

A.~de Bari
\\{\it 
Sezione INFN Pavia and Dipartimento di Fisica, Universit\`{a} di Pavia, Via Agostino Bassi 6, Pavia, Italy
}\\

D.~Orestano, L.~Tortora
\\{\it
Sezione INFN Roma Tre and Dipartimento di Matematica e Fisica, Universit\`{a} Roma Tre, Via della Vasca Navale 84, Roma, Italy
}\\

Y.~Kuno, H.~Sakamoto\footnote{Current address RIKEN 2-1 Hirosawa, Wako, Saitama, Japan}, A.~Sato
\\{\it
Osaka University, Graduate School of Science, Department of Physics, 1-1 Machikaneyamacho, Toyonaka, Osaka, Japan
}\\

S.~Ishimoto
\\{\it
High Energy Accelerator Research Organization (KEK), Institute of Particle and Nuclear Studies, Tsukuba, Ibaraki, Japan
}\\

M.~Chung, C.~K.~Sung
\\{\it 
Department of Physics, UNIST, 50 UNIST-gil, Ulsan, South Korea
}\\

F.~Filthaut\footnote{Also at Radboud University, Houtlaan 4, Nijmegen, Netherlands}
\\{\it
Nikhef, Science Park 105, Amsterdam, Netherlands
}\\

M.~Fedorov
\\{\it
Radboud University, Houtlaan 4, Nijmegen, Netherlands
}\\

D.~Jokovic, D.~Maletic, M.~Savic
\\{\it
Institute of Physics, University of Belgrade, Serbia
}\\

%\newpage

N.~Jovancevic, J.~Nikolov
\\{\it
Faculty of Sciences, University of Novi Sad, Trg Dositeja Obradovi\'{c}a 3, Novi Sad, Serbia
}\\

M. ~Vretenar, S.~Ramberger
\\{\it
CERN, Esplanade des Particules 1, Geneva, Switzerland
}\\

R.~Asfandiyarov, A.~Blondel, F.~Drielsma\footnote{Current address SLAC National Accelerator Laboratory, 2575 Sand Hill Road, Menlo Park, California, USA}, Y.~Karadzhov 
\\{\it
DPNC, Section de Physique, Universit\'e de Gen\`eve, 24 Quai Ernest-Ansermet, Geneva, Switzerland
}\\

G.~Charnley, N.~Collomb,  K.~Dumbell, A.~Gallagher, A.~Grant, S.~Griffiths,  T.~Hartnett, B.~Martlew, 
A.~Moss, A.~Muir, I.~Mullacrane, A.~Oates, P.~Owens, G.~Stokes, P.~Warburton, C.~White
\\{\it
STFC Daresbury Laboratory, Keckwick Ln, Daresbury, Cheshire, UK
}\\

D.~Adams,   V.~Bayliss, J.~Boehm, T.~W.~Bradshaw, C.~Brown\footnote{Also at College of Engineering, Design and Physical Sciences, Brunel University, Kingston Lane, Uxbridge, UK}, M.~Courthold,  J.~Govans, M.~Hills, J.-B.~Lagrange, C.~Macwaters, A.~Nichols, R.~Preece, S.~Ricciardi, C.~Rogers, T.~Stanley, J.~Tarrant,  
M.~Tucker, S.~Watson\footnote{Current address ATC, Royal Observatory Edinburgh, Blackford Hill, Edinburgh, UK},A.~Wilson
\\{\it
STFC Rutherford Appleton Laboratory, Harwell Campus, Didcot, UK
}\\

R.~Bayes\footnote{Current address Laurentian University, 935 Ramsey Lake Road, Sudbury, Ontario, Canada},  J.~C.~Nugent, F.~J.~P.~Soler
\\{\it
School of Physics and Astronomy, Kelvin Building, University of Glasgow, Glasgow, UK
}\\

R.~Gamet, P.~Cooke
\\{\it
Department of Physics, University of Liverpool, Oxford St, Liverpool, UK
}\\

V.~J.~Blackmore, D.~Colling, A.~Dobbs\footnote{Current address OPERA Simulation Software, Network House, Langford Locks, Kidlington, UK}, P.~Dornan, P.~Franchini, C.~Hunt\footnote{Current address CERN, Esplanade des Particules 1, Geneva, Switzerland}, P.~B.~Jurj, A.~Kurup, K.~Long, J.~Martyniak,  S.~Middleton\footnote{Current address School of Physics and Astronomy, University of Manchester, Oxford Road, Manchester, UK}, J.~Pasternak, M.~A.~Uchida\footnote{Current address Rutherford Building, Cavendish Laboratory, JJ Thomson Avenue, Cambridge, UK}
\\{\it
Department of Physics, Blackett Laboratory, Imperial College London, Exhibition Road, London, UK
}\\

J.~H.~Cobb
\\{\it
Department of Physics, University of Oxford, Denys Wilkinson Building, Keble Rd, Oxford, UK
}\\

C.~N.~Booth, P.~Hodgson, J.~Langlands, E.~Overton\footnote{Current address Arm, City Gate, 8 St Mary's Gate, Sheffield, UK}, V.~Pec,  P.~J.~Smith, S.~Wilbur
\\{\it
Department of Physics and Astronomy, University of Sheffield, Hounsfield Rd, Sheffield, UK
}\\

G.~T.~Chatzitheodoridis\footnote{Also at School of Physics and Astronomy, Kelvin Building, University of Glasgow, Glasgow, UK}$^,$\footnotemark, A.~J.~Dick\footnotemark[\value{footnote}],  K.~Ronald\footnotemark[\value{footnote}], C.~G.~Whyte\footnotemark[\value{footnote}], A.~R.~Young\footnotemark[\value{footnote}]
\footnotetext{Also at Cockcroft Institute, Daresbury Laboratory, Sci-Tech Daresbury, Keckwick Ln, Daresbury, Warrington, UK}
\\{\it
SUPA and the Department of Physics, University of Strathclyde, 107 Rottenrow, Glasgow, UK
}\\

S.~Boyd,  J.~R.~Greis\footnote{Current address TNG Technology Consulting, Beta-Strasse 13A, Unterf\"{o}hring, Germany}, T.~Lord, C.~Pidcott\footnote{Current address Department of Physics and Astronomy, University of Sheffield, Hounsfield Rd, Sheffield, UK}, I.~Taylor\footnote{Current address Defence Science and Technology Laboratory, Porton Down, Salisbury, UK}
\\{\it
Department of Physics, University of Warwick, Gibbet Hill Road, Coventry, U
}\\

M.~Ellis\footnote{Current address Macquarie Group, 50 Martin Place, Sydney, Australia}, R.B.S.~Gardener\footnote{Current address Inawisdom, Columba House, Adastral park, Martlesham, Ipswich, UK}, P.~Kyberd, J.~J.~Nebrensky
\\{\it
College of Engineering, Design and Physical Sciences, Brunel University, Kingston Lane, Uxbridge, UK
}\\

M.~Palmer, H.~Witte
\\{\it
Brookhaven National Laboratory, 98 Rochester St, Upton, New York, USA
}\\

\newcounter{FNEuclid}
D.~Adey\footnote{Current address Institute of High Energy Physics, Chinese Academy of Sciences, 19 Yuquan Rd, Shijingshan District, Beijing, China}, A.~D.~Bross, D.~Bowring, P.~Hanlet, A.~Liu\footnotemark\footnotetext{Current address Euclid Techlabs, 367 Remington Blvd, Bolingbrook, Illinois, USA}\setcounter{FNEuclid}{\value{footnote}}, D.~Neuffer, M.~Popovic, P.~Rubinov
\\{\it
Fermilab, Kirk Rd and Pine St, Batavia, Illinois, USA
}\\

A.~DeMello, S.~Gourlay, A.~Lambert, D.~Li, T.~Luo, S.~Prestemon,  S.~Virostek
\\{\it
Lawrence Berkeley National Laboratory, 1 Cyclotron Rd, Berkeley, California, USA
}\\

B.~Freemire\footnotemark[\value{FNEuclid}], D.~M.~Kaplan, T.~A.~Mohayai\footnote{Current address Fermilab, Kirk Rd and Pine St, Batavia, Illinois, USA}, D.~Rajaram\footnote{Current address KLA, 2350 Green Rd, Ann Arbor, Michigan, USA}, P.~Snopok, Y.~Torun
\\{\it
Illinois Institute of Technology, 10 West 35th St, Chicago, Illinois, USA
}\\

L.~M.~Cremaldi, D.~A.~Sanders, D.~J.~Summers\footnote{Deceased}
\\{\it
University of Mississippi, University Ave, Oxford, Mississippi, USA
}\\

L.~R.~Coney\footnote{Current address European Spallation Source ERIC, Partikelgatan 2, Lund, Sweden}, G.~G.~Hanson, C.~Heidt\footnote{Current address Swish Analytics, Oakland, California, USA}
\\{\it
University of California, 900 University Ave, Riverside, California, USA
}\\

\end{document}